\documentclass[%
 amsmath,amssymb,
 aps,
 pra,
]{revtex4-2}

\usepackage{graphicx}% Include figure files
\usepackage{dcolumn}% Align table columns on decimal point
\usepackage{bm}% bold math
\usepackage{hyperref}% add hypertext capabilities
\newcommand{\figref}[1]{Fig.~\ref{#1}}
\newcommand{\unit}[1]{~\mathrm{#1}}
\newcommand{\dif}{\mathrm{d}}%ok????

\begin{document}

%\preprint{APS/123-QED}

\title{Active colloid with externally induced periodic bipolar motility \\and its cooperative motion}% Force line breaks with \\

\author{Airi N. Kato$^{1,2}$}\email{airi@ucas.ac.cn}\author{Kazumasa A. Takeuchi$^{1}$}\email{kat@kaztake.org}\author{Masaki Sano$^{3}$}\email{sano.masaki@sjtu.edu.cn}
\affiliation{$^{1}$Department of Physics, The University of Tokyo, 7-3-1 Hongo, Bunkyo-ku,Tokyo 113-0033, Japan.}
\affiliation{$^{2}$Wenzhou Institute, University of Chinese Academy of Sciences, Wenzhou, Zhejiang 325001, China}
%\affiliation{Department of Physics, The University of Tokyo, 7-3-1 Hongo, Bunkyo-ku,Tokyo 113-0033, Japan.}
\affiliation{
 $^{3}$Institute of Natural Sciences, School of Physics and Astronomy, Shanghai Jiao Tong University, Shanghai 200240, China.
}%

\date{\today}% It is always \today, today,

\begin{abstract}
Active matter physics has been developed with various types of self-propelled particles, including those with polar and bipolar motility and beyond. However, the bipolar motions experimentally realized so far have been either random along the axis or periodic at intrinsic frequencies. Here we report another kind of bipolar active particles, whose periodic bipolar self-propulsion is set externally at a controllable frequency. We used Quincke rollers---dielectric particles suspended in a conducting liquid driven by an electric field---under an AC electric field instead of the usually used DC field. Reciprocating motion of a single particle at the external frequency was observed experimentally and characterized theoretically as stable periodic motion. Experimentally, we observed not only the reciprocating motion but also non-trivial active Brownian particle (ABP)-like persistent motion in a long time scale. This resulted in a Lorentzian spectrum around zero frequency, which is not accounted for by a simple extension of the conventional model of Quincke rollers to the AC field. It was found that ABP-like motion can be reproduced by considering the top-bottom asymmetry in the experimental system. Moreover, we found a rotational diffusion coefficient much larger than the thermal one, as also reported in previous experiments, which may have resulted from roughness of the electrode surface.  We also found self-organized formation of small clusters, such as doublets and triplets, and characterized cooperative motion of particles therein. The AC Quincke rollers reported here may serve as a model experimental system of bipolar active matter, which appears to deserve further investigations.
\end{abstract}

\maketitle

%\tableofcontents

\section{Introduction}
Active matter is a class of intrinsically non-equilibrium systems, usually consisting of a collection of self-propelled particles (SPPs) that move according to the individual polarity or axis of motility \cite{Marchetti2013r,Bechinger.etal-RMP2016,Doostmohammadi.etal-NC2018,Chate-ARCMP2019}.
Characteristic features of active matter that do not appear in equilibrium systems, such as collective motions ({\it e.g.}, ordering \cite{Vicsek1995,Marchetti2013r,Chate-ARCMP2019} and dynamic clustering \cite{Theurkauff2012,Palacci2013,Buttinoni2013,Ginot2018,Bechinger.etal-RMP2016}) and giant number fluctuations \cite{Toner.Tu-PRL1995,Ramaswamy.etal-EL2003,Narayan2007,Marchetti2013r,Chate-ARCMP2019}, have been a target of intensive studies for decades.
These include experimental studies using systems from microscopic to macroscopic scales, and from biological to artificial systems. Examples of artificial experimental systems are shaken rods and disks \cite{Narayan2007,Deseigne2010}, Janus particles \cite{Paxton.etal-JACS2004,Wang2006,Gangwal2008,Jiang2010,Theurkauff2012,Palacci2013,Buttinoni2013,Nishiguchi.Sano-PRE2015,Mano.etal-PNAS2017,Ginot2018,Iwasawa.etal-a2020}, Quincke rollers \cite{Bricard2013,Bricard2015,Karani2019,Zhang2021}, floating droplets \cite{Nagai2005,Ebata2015}, and so on (see \cite{Bechinger.etal-RMP2016} for a review).

One of the main advantages of using artificial self-propelled particles is controllability. Typically, self-propulsion speed can be controlled externally within some range; for example, speed of Janus particles driven by induced-charge electrophoresis \cite{Gangwal2008,Nishiguchi.Sano-PRE2015,Mano.etal-PNAS2017,Iwasawa.etal-a2020} and that of Quincke rollers \cite{Bricard2013,Bricard2015} can be controlled by the strength of the electric field.
In such systems driven by an external field, active and passive states can also be switched by turning the field on and off \cite{Lozano2018,Qian2013,Karani2019} or otherwise \cite{Mano.etal-PNAS2017}, leading to motions reminiscent of run-and-tumbling of {\it E.coli.} \cite{Berg1972}.
In these cases, the motion is unidirectional, \textit{i.e.}, particles are self-propelled in the direction of the polarity while the field is on.
In contrast, some systems such as shaken rods \cite{Narayan2007} exhibit bipolar motion by stochastic reversal of self-propelling direction.
Together with biological examples such as {\it Myxococcus xanthus} \cite{Wu2009} and neural progenitor cells \cite{Kawaguchi.etal-N2017}, these systems have been studied as examples of active nematics \cite{Doostmohammadi.etal-NC2018}.
There are also examples of bipolar active particles with  periodic reversal of self-propelling direction, such as self-propelled Belousov-Zhabotinsky droplets \cite{Thutupalli2013}, a surface-wave driven droplet \cite{Ebata2015}, and Quincke rollers under a strong DC electric field \cite{Zhang2021}.
However, in all these examples, the periodic reversal takes place at a frequency intrinsic to each system, which is therefore uncontrollable.

\begin{figure}[t]
\centering
  \includegraphics[width=8.9cm]{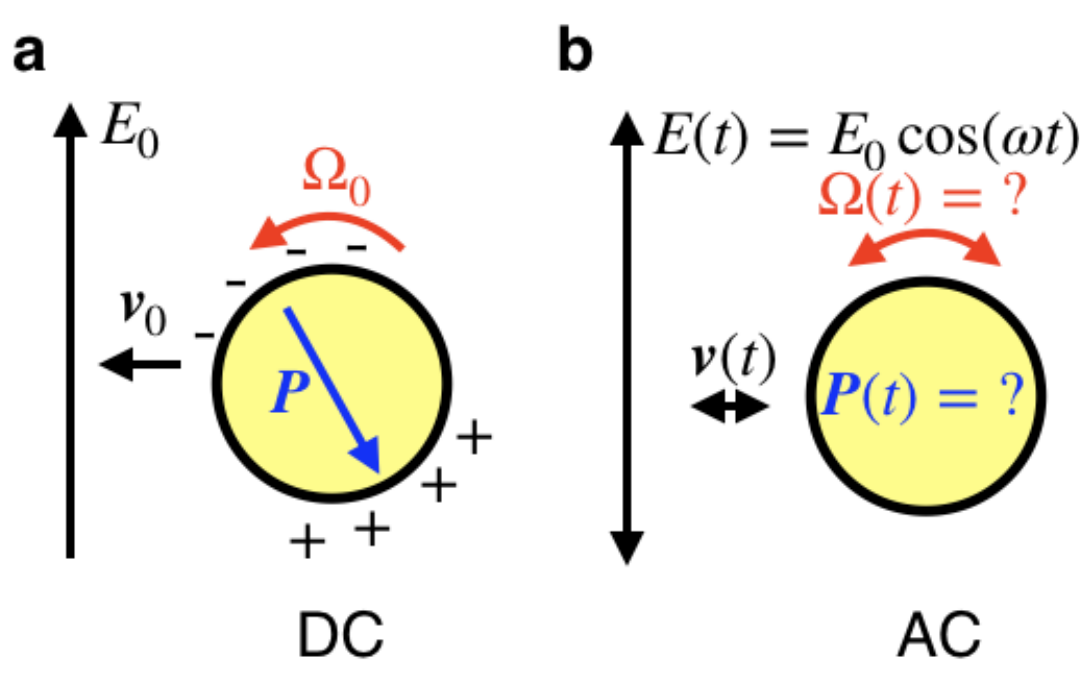}
  \caption{Schematics of a Quincke roller under a DC field (a) and an AC field (b). (a) Under a DC field, a constant polarization is formed and inclined in steady state, leading to a constant self-propelling velocity. (b) Under an AC field, the polarization and the velocity become time-dependent. We found periodic bipolar motility at the frequency of the external field in this case.}
  \label{fig:ponchie}
\end{figure}

Here we report that Quincke rollers, driven by an AC electric field instead of the usually used DC field, constitute an experimental system of bipolar active particles with periodic reversal of self-propelling direction at an externally set frequency (Fig.~\ref{fig:ponchie}).
We call our system ``AC Quincke rollers'' afterwards to distinguish from the original DC Quincke rollers.
A DC Quincke roller consists of a dielectric sphere immersed in a conducting liquid under a constant DC electric field ${\bf E}_0$. The particle rolls on the electrode by the Quincke effect \cite{Quincke} as shown in Fig.~\ref{fig:ponchie}a.
The Quincke effect occurs under the condition that the relaxation time of ions in liquid, $\tau_l:= \epsilon_l/\sigma_l$, is shorter than that on the surface of the particle, $\tau_p:= \epsilon_p/\sigma_p$.
Here, $\epsilon_{l,p}$ is the dielectric constant of the liquid and the particle, respectively, and $\sigma_{l,p}$ is the electric conductivity of the liquid and the particle, respectively.
When the electric field ${\bf E}_0$ is applied, charges accumulate on the surface of the particle, which makes an effective polarization ${\bf P}$ antiparallel to the field (Fig.~\ref{fig:ponchie}a).
This configuration is unstable because a small perturbation of the polarization perpendicular to the field is amplified by the electric torque ${\bf P}\times {\bf E}_0$.
Therefore, if this torque surpasses the viscous restitution torque, the particle starts to rotate spontaneously.
As a result, the particle rolls on the substrate electrode (Fig.~\ref{fig:ponchie}a), leading to unidirectional self-propelling motion in the two-dimensional plane. The direction of motion is selected by spontaneous symmetry breaking.
Note that the particle itself has no polarity without the external electric field.
It is the polarization ${\bf P}$ that determines the direction of motion (polarity), which exists only when the external field is applied.
The fact that the polarity is acquired only \textit{a posteriori} under the external field is the key to realize the externally induced periodic bipolar motion realized by our AC Quincke rollers.

In this paper, we focus on the behavior of the AC Quincke rollers in a dilute suspension.
First we describe the experimental setup (Sec.~\ref{sec:Setup}) and show experimental results on the motion of a single particle (Sec.~\ref{sec:singleE}).
We found that a single particle exhibits periodic reciprocating motion with active Brownian particle (ABP)-like persistent net motion.
Next, the theoretical description of a Quincke roller under a DC field was generalized to that under an AC field (Sec.~\ref{sec:singleT}).
It was found that the reciprocating motion at the frequency of the external field corresponds to a limit cycle of the equations of motion.
This supports our observation that AC Quincke rollers are a realization of controllable SPPs with stable reciprocating motions at the frequency of the external field.
The ABP-like motion was reproduced by incorporating the asymmetry caused by the existence of the lower electrode.
Then we focus on motion of two- and three-particle clusters (doublet and triplet) which were spontaneously formed and maintained for long periods of time (at least $100$--$1000$ periods) (Sec.~\ref{sec:MultiParticle}).
Characteristic correlations in particle motions are revealed and interpreted in terms of particle interactions.
We believe that the AC Quincke rollers reported here may provide a new perspective on active matter systems with periodic bipolar self-propulsion, through nontrivial consequences that such time dependence brings.

\begin{figure}[t]
\centering
  \includegraphics[width=15.1cm]{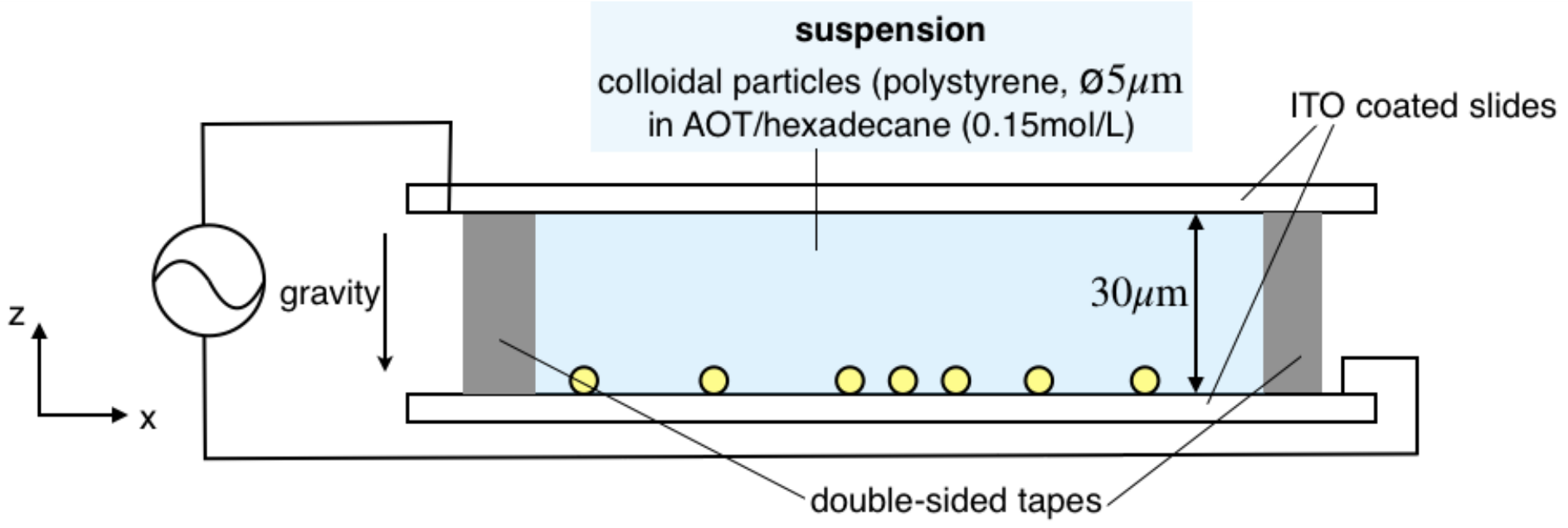}
  \caption{Experimental setup. A suspension of polystyrene spheres including surfactant (AOT) is sandwiched between two transparent electrodes. The particles roll on the lower substrate under an electric field, moving thereby on the quasi-two-dimensional plane.}
  \label{fig:setup}
\end{figure}

\section{Experimental setup and methods}  \label{sec:Setup}

Quincke rollers under an AC electric field were experimentally realized as follows.
Following Bricard {\it et al.} \cite{Bricard2013,Bricard2015} on the DC Quincker rollers, we used polystyrene spherical colloids (Thermo Scientific GS0500, radius $2.5\unit{\mu m}$) dispersed in $0.15\unit{mol/L}$ of AOT/hexadecane solution.
As shown in \figref{fig:setup}, the suspension was sandwiched between two transparent electrodes (glass plate coated with indium-tin oxide (ITO) film, Mitsuru Optical Co. Ltd.).
The electrodes were spaced by $30\unit{\mu m}$ thick double-sided polyethylene terephthalate tapes (NITTO, No.5603).
The AC electric field was applied by a function generator (Agilent, 33220A) through an amplifier (NF, TA-120). \footnote{Applying AC electric fields has the additional advantage of not damaging electrodes, in contrast applying DC electric field damages electrodes in about 10 minutes due to chemical reactions on the surface.}
The particles were observed by bright-field microscopy using an inverted microscope (Olympus, IX70) with an x40 objective lens (Olympus, LUCPLFLN40X, NA $0.60$) unless otherwise stipulated,
% or x10 objective lens (Olympus, UPlanFL10X, NA$0.30$)
and captured by a high-speed camera (Photron, Fastcam mini AX) at $3000\unit{fps}$ for $300\unit{Hz}$ and  $2500\unit{fps}$ for $250\unit{Hz}$. The image acquisition was synchronized by the trigger signal from the function generator.
Tracking was conducted using Particle Tracker 2D/3D \cite{Sbalzarini2005}, a plugin of ImageJ.
When an AC electric field was applied, we observed approximately periodic motion after a short transient time. We therefore conducted all measurements after waiting long enough, specifically at least 5 seconds after the start of the voltage application.
In this paper, unless otherwise stipulated, we fixed the root-mean-square amplitude and the frequency of the AC applied voltage at $150\unit{V}$ and $300\unit{Hz}$ or $250\unit{Hz}$, respectively.
%, which corresponds to the period of $T=1/300\unit{s}$ or $T=1/250\unit{s}$.

\section{Experimental results on single AC Quincke rollers}\label{sec:singleE}

\begin{figure*}[t]
\centering
  \includegraphics[width=18.1cm]{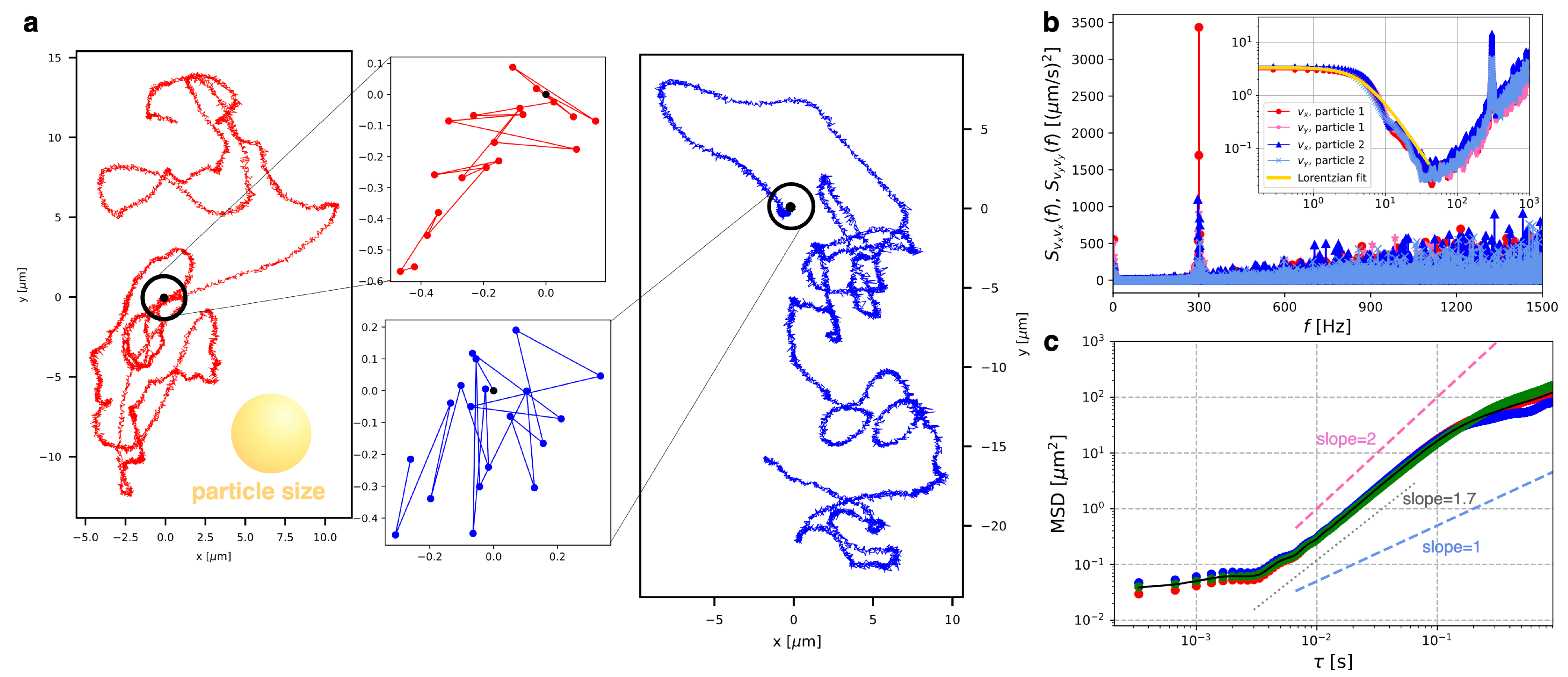}
  \caption{Single-particle behavior ($300\unit{Hz}, 150\unit{V}$). (a) Two independent trajectories (over $1091$~periods $= 3.639\unit{s}$). The subpanels show the positions of the particles recorded every half period for the first $10$~periods. The initial positions are depicted by the black dots. The two particles were captured simultaneously, with the distance varying in the range of $177$--$198\unit{\mu m}$. (b) Energy spectral densities. The particles 1 and 2 in the legend correspond to the red and blue trajectories in (a), respectively. The inset shows an enlargement of a low-frequency region. To reduce statistical fluctuations at a low-frequency region, the single whole time series was divided into $30$ segments (each segment contains $36$ periods), then the power spectra were averaged. The yellow solid line shows the result of the fitting to the average of the four data by the Lorentzian spectrum $v_0^2 \frac{2D_{\theta}}{D_{\theta}^2+(2\pi f)^2}$ in Eq.~\eqref{eq:spectrum}. The data for $f<40\unit{Hz}$ was used for fitting. (c) The MSDs of three isolated particles (red, blue, and green dots) and their average (black solid line). The dashed lines are guides for eyes with the indicated slopes.}
  \label{fig:single}
\end{figure*}

This section reports experimental results on single particle motion.
Here we used isolated particles, or more specifically, particles separated more than $100\unit{\mu m}$ from others.
Typical trajectories of isolated particles at $300\unit{Hz}$ are shown in \figref{fig:single}a.
In short time scales, the observed particles exhibited reciprocating motion at the frequency of the external field, with some extent of randomness in the direction and amplitude of the reciprocating motion (Fig. \ref{fig:single}a, inset).
By contrast, in time scales much longer than the reciprocation period, we observed persistent motion whose direction changed gradually (see Videos~S1 and S2 in ESI\dag).
Using each component of the single particle velocity ${\bf v}=(v_x,v_y)$, we computed the energy spectral density %$S_{vv}(f):=|\hat{v}(f)|^2$, where $\hat{v}(f):=1/N\sum_{t=0}^{N-1}v(t)e^{-2\pi i ft/N}$.
\begin{equation}
    S_{vv}(f):=|\hat{v}(f)|^2,\qquad \hat{v}(f):= \frac{1}{N}\sum_{k=0}^{N-1}v(t_k)e^{-2\pi i f t_k},
\end{equation}
with $v(t)=v_x(t)$ or $v_y(t)$, where $t_k$ is the time at the $k$th video frame and $N$ is the total frame number.
The results are shown in \figref{fig:single}b. To reduce statistical fluctuations in a low-frequency region, we computed the power spectra in the inset of \figref{fig:single}b by segmenting the single whole time series into 30 segments and then averaging them.
This shows that the velocities had the periodicity at the external frequency, $300\unit{Hz}$.
The spectra also showed another signal in a low-frequency region ($\lesssim 10\unit{Hz}$) as shown in the inset of \figref{fig:single}b, which corresponds to the slowly changing, persistent motion we observed. This low-frequency mode was also observed for 200Hz and 250Hz (Fig.~S2, ESI\dag). In fact, the spectral density in the low-frequency region is Lorentzian, which is expected from a simple one-particle model with a sinusoidal and a nonzero constant self-propelling velocity $v(t)=v_0+v_1\cos\omega t$ (see Supplemental Notes, ESI\dag), and also seen for the ABP \cite{Jiang2010,Squarcini2021}. More precisely, the spectral density $S(f)$ for the model with $v(t)=v_0+v_1\cos \omega t$ is
\begin{equation}
  S(f)=v_0^2 \frac{2D_{\theta}}{D_{\theta}^2+(2\pi f)^2}+\frac{v_1^2}{2}\left(\frac{D_{\theta}}{D_{\theta}^2+(\omega+2\pi f)^2}+\frac{D_{\theta}}{D_{\theta}^2+(\omega-2\pi f)^2}\right),\label{eq:spectrum}
\end{equation}
where $D_\theta$ is the rotational diffusion coefficient. This coefficient was estimated at $D_\theta=31.4\pm0.6 \unit{s}^{-1}$ ($D_\theta^{-1}\sim 0.03\unit{s}$) by the Lorentzian (the first term of Eq.~\eqref{eq:spectrum}) fitting to the average of the four spectra in the inset of \figref{fig:single}b for $<40\unit{Hz}$; here the fitting was done with the peak value of the Lorentzian fixed at $S(0)$. The velocities $v_0$ and $v_1$ were estimated at $v_0=7.3\pm 0.4\unit{\mu m/s}$ and $v_1=24\pm 8 \unit{\mu m/s}$ by the peaks at $0$ and $300\unit{Hz}$. Note that this effective rotational diffusion coefficient $D_\theta$ is four orders of magnitude larger than the thermal rotational diffusion coefficient $D_{\theta}=k_{\mathrm{B}} T/(8\pi \eta a^3 )\sim 10^{-3}\unit{s}^{-1}$, similarly to the case of Bricard {\it et al.} \cite{Bricard2015}. They reported $D_\theta^{-1}\sim 0.31\unit{s}$ in the DC case. In contrast to these Quincke experiments, such a considerable deviation between the measured rotational diffusion coefficient of a spherical particle and the thermal value was not observed in experiments of thermally activated Janus particles\cite{Jiang2010}, which did not use an electrode.
Therefore, this athermal rotational diffusion may be originated from properties of the electrode, such as its surface roughness. To inspect this possibility, we measured surface roughness of the ITO film by using an atomic force microscope (AFM) and discussed possible effects of quenched noise originated from the surface roughness in the Supplemental Notes (ESI\dag). The origin of the DC component of the velocity will be discussed in Sec.~\ref{sec:singleT}.

To characterize the dynamics of the AC Quincke rollers, we also computed their mean square displacements (MSDs):
\begin{equation}
    \mathrm{MSD}(\tau)=\frac{1}{N-m}\sum_{k=1}^{N-m}({\bf x}(t_{k+m})-{\bf x}(t_{k}))^2,
\end{equation}
where $\tau = (m/3000)\unit{s}$ is the lag time, and ${\bf x}(t_k)$ is the particle position at time $t_k$.
The results for three isolated particles are shown in \figref{fig:single}c.
The MSDs oscillate with the period $T = (1/300)\unit{s}$ of the applied field and grow algebraically over a duration longer than $20T$ with a superdiffusive exponent $1.70 \pm 0.07$ as shown by the black dashed line in \figref{fig:single}c.
This estimate was obtained by using the data points taken at every $10$ frames ($1$ period) in the range $T\leq\tau\leq 20T$, and the 95\% confidence interval was given.
We consider that this superdiffusion is associated with the low-frequency mode that manifested itself in the long-time persistent motion.
After this superdiffusive behavior, the MSD exponent crossovers to the diffusive one.
This crossover from the nearly ballistic behavior to the diffusive one is another aspect similar to the ABP, in addition to the aforementioned Lorentzian around $0\unit{Hz}$ in the spectral densities characterized by the effective rotational diffusion.

%On the other hand, the exponent in long-time was $0.726\pm 0.004$ (the 95\% confidence interval for the fitting in the range $0.2\unit{s}\leq\tau\leq 1\unit{s}$), subdiffusive?

\section{Theory on single AC Quincke rollers}\label{sec:singleT}

\begin{figure*}[t]
\centering
  \includegraphics[width=18.1cm]{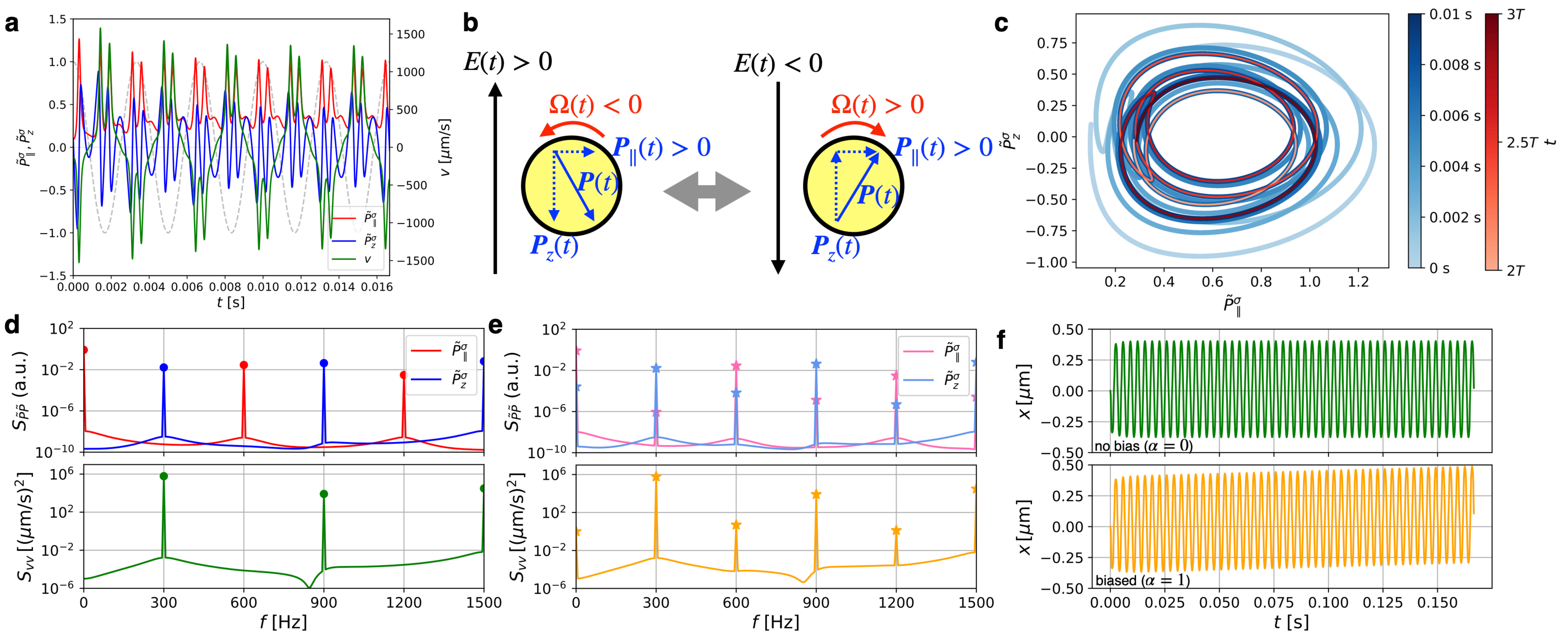}
  \caption{Single-particle behavior in the theoretical model ($300\unit{Hz},\, 150\unit{V}$). (a) Time evolution of the non-dimensional polarization $\tilde{P}_{\parallel}^\sigma$ (red) and $\tilde{P}_z^\sigma$ (blue), and the particle velocity $v$ (green). We set $\frac{a \mu_t}{\mu_r \tau_{\mathrm{MW}}}=82.1\unit{\mu m/s}$. (b) Sketch of the reciprocating rolling mechanism of an AC Quincke roller. Note that the sign of $\tilde{P}_{\parallel}^\sigma$ is kept the same. (c) A trajectory of  $(\tilde{P}_{\parallel}^\sigma, \tilde{P}_z^\sigma)$ from a transient to the limit cycle. The red line shows the periodic cycle over one period. The color gradient shows the time change. (d) The autospectra of the polarization (top) and the velocity (bottom) from the data in (a) in the periodic state. (e) The autospectra of the polarization (top) and the velocity (bottom) in the biased model ($\alpha=1$). (f) The positions of a particle $x(t)$ with no bias ($\alpha=0$) and with a bias ($\alpha=1$).}
\label{fig:singleth}
\end{figure*}

To gain insights on the single particle dynamics observed in the previous section, here we provide a theoretical description of single AC Quincke rollers, by extending the theory for DC Quincke rollers formulated by Bricard \textit{et al.} \cite{Bricard2013}.
In the DC case, the locomotion of a single Quincke roller of radius $a$ under an electric field ${\bf E}(t) = {\bf E}_0$ is described by its angular momentum $\pmb{\Omega}(t)$, which obeys the equation of motion
\begin{equation}\label{eq:eom}
I\frac{\dif\pmb{\Omega}}{\dif t}=\frac{\epsilon_l}{\epsilon_0}({\bf P}\times {\bf E}_0) - \mu_r^{-1} \pmb{\Omega}
\end{equation}
with the rotational inertia $I$, the dielectric constant of vacuum $\epsilon_0$, and the rotational mobility $\mu_r$ expected to be affected by lubrication \cite{Goldman1967,Goldman1967-2,Oneill1967,Liu2010,Bricard2013}.

The polarization ${\bf P}(t)$ evolves by the following electrodynamic equation \cite{Melcher1969,Pannacci2007,Bricard2013}:
\begin{equation}
\frac{\dif{\bf P}}{\dif t}+\frac{1}{\tau_{\mathrm{MW}}}{\bf P}=-\frac{2 \pi \epsilon_0 a^3}{\tau_{\mathrm{MW}}}{\bf E}_0+\pmb{\Omega}\times ({\bf P}-4\pi\epsilon_0 a^3 \chi^{\infty}{\bf E}_0), \label{eq:electrodynamics}
\end{equation}
with $\chi^{\infty}=\frac{\epsilon_p-\epsilon_l}{\epsilon_p+2\epsilon_l}$ and the Maxwell-Wagner time $\tau_{\mathrm{MW}}=\frac{\epsilon_p+2\epsilon_l}{2\sigma_l}$.
Since the inertia term is negligible ($\mu_rI\sim 10^{-8}\unit{s}\ll \tau_{\mathrm{MW}}$ in our system), we have $\pmb{\Omega}(t) = \mu_r\frac{\epsilon_l}{\epsilon_0}({\bf P}\times {\bf E}_0)$.
At this point, it is convenient \cite{Bricard2013} to recognize that ${\bf P}(t)$ consists of the following two contributions: ${\bf P}(t) = {\bf P}^\epsilon(t) + {\bf P}^\sigma(t)$, where ${\bf P}^\epsilon(t) = 4\pi \epsilon_0 a^3 \chi^\infty {\bf E}_0$ is the dielectric contribution due to the permittivity discontinuity at the interface, and the rest ${\bf P}^\sigma(t)$ is due to the surface charge transport.
The charge transport contribution ${\bf P}^\sigma(t)$ can be further decomposed into the $z$-component $P_z^\sigma(t)$ and the component parallel to the surface, ${\bf P}_\parallel^\sigma(t)$.
With this and $\pmb{\Omega}(t) = \mu_r\frac{\epsilon_l}{\epsilon_0}({\bf P}\times {\bf E}_0)$, the time evolution equations for $P_z^\sigma(t)$ and ${\bf P}_\parallel^\sigma(t)$ read
\begin{align}
    &\frac{\dif {P_z^\sigma}}{\dif t}+\frac{1}{\tau_{\mathrm{MW}}}{P_z^\sigma} = -\frac{4\pi\epsilon_0 a^3}{\tau_{\mathrm{MW}}}\left(\chi^\infty + \frac{1}{2}\right)E_0  + \mu_r\frac{\epsilon_l}{\epsilon_0}E_0 (P_\parallel^\sigma)^2,  \label{eq:P1} \\
    &\frac{\dif {{\bf P}_\parallel^\sigma}}{\dif t}+\frac{1}{\tau_{\mathrm{MW}}}{{\bf P}_\parallel^\sigma} = - \mu_r\frac{\epsilon_l}{\epsilon_0} E_0 P_z^\sigma {\bf P}_\parallel^\sigma.  \label{eq:P2}
\end{align}
%\begin{eqnarray}
%    \frac{\dif {\bf P}}{\dif t}+\frac{1}{\tau_{\mathrm{MW}}}{\bf P}    =-\frac{2\pi\epsilon a^3}{\tau_{\mathrm{MW}}}{\bf E}_0+\mu_r\frac{\epsilon_l}{\epsilon_0}\biggl[|{\bf P}|^2{\bf E}_0-({\bf E}_0\cdot {\bf P}){\bf P}\nonumber\\
%    -4\pi\epsilon_0 a^3\chi^{\infty} ({\bf P}\cdot {\bf E}_0){\bf E}_0+4\pi\epsilon_0 a^3\chi^{\infty}|{\bf E}_0|^2 {\bf P}\biggr].\label{eq:P}
%\end{eqnarray}
For a strong enough field $E_0$, this set of the equations has a stable fixed point corresponding to the motion at a constant translation velocity.
It is given by
\begin{equation}\label{eq:vconstantdc}
{\bf v} = -\frac{\epsilon_l}{\epsilon_0}a\mu_tE_0{\bf P}_{\parallel}^\sigma,
\end{equation}
with the cross mobility coefficient $\mu_t$ that relates the electric torque to the translational velocity \cite{Bricard2013}.

Now we generalize this formulation to deal with an oscillating electric field, ${\bf E}(t) = {\bf E}_0 \cos\omega t$.
%In fact, Eq.~\eqref{eq:vconstantdc} with ${\bf E}_0$ replaced by ${\bf E}_0 \cos\omega t$ already suggests that the velocity ${\bf v}(t)$ can oscillate at the frequency of the external field, if the parallel component of the polarization ${\bf P}_{\parallel}(t)$ has a non-vanishing time average.
Replacing ${\bf E}_0$ in Eqs.~\eqref{eq:eom} and \eqref{eq:electrodynamics} with ${\bf E}_0 \cos\omega t$ and using ${\bf P}(t) = {\bf P}^\epsilon(t) + {\bf P}^\sigma(t)$ with ${\bf P}^\epsilon(t) = 4\pi \epsilon_0 a^3 \chi^\infty {\bf E}_0 \cos\omega t$, under the assumption that dielectric polarization takes place fast enough compared to $T=2\pi/\omega$ (typical dielectric polarization response is considered to be instantaneous, much faster than our external frequency: $300\unit{Hz}$), we obtain the following time evolution equations:
\begin{align}
    \frac{\dif {P_z^\sigma}}{\dif t}+\frac{1}{\tau_{\mathrm{MW}}}{P_z^\sigma} &= -4\pi\epsilon_0 a^3 E_0 \left(\frac{\chi^\infty + 1/2}{\tau_{\mathrm{MW}}} \cos\omega t - \chi^\infty \omega\sin\omega t \right) \notag \\
    & \qquad + \mu_r\frac{\epsilon_l}{\epsilon_0} E_0 (P_\parallel^\sigma)^2 \cos\omega t,  \label{eq:P-AC1z0} \\
    \frac{\dif {{\bf P}_\parallel^\sigma}}{\dif t}+\frac{1}{\tau_{\mathrm{MW}}}{{\bf P}_\parallel^\sigma} &= - \mu_r\frac{\epsilon_l}{\epsilon_0} E_0 P_z^\sigma {\bf P}_\parallel^\sigma \cos\omega t.  \label{eq:P-AC1x0}
\end{align}
The rolling velocity of the particle is
\begin{equation}\label{eq:v_AC}
{\bf v}(t)=-\frac{\epsilon_l}{\epsilon_0}a\mu_t E_0 {\bf P}_{\parallel}^\sigma (t) \cos \omega t.
\end{equation}
Now, we introduce the following non-dimensionalized quantities: $\tilde{t}:=t/\tau_{\mathrm{MW}}$, $\tilde{\omega}:=\omega\tau_{\mathrm{MW}}$, $\tilde{\bf P}:=\frac{\bf P}{\varepsilon_0 E_0 a^3}$, and $A:=\mu_r \varepsilon_l E_0^2 a^3\tau_{\mathrm{MW}}$. The rolling velocity, along the axis of motion is
\begin{equation}\label{eq:v_AC2}
v(t)=-\frac{a\mu_t}{\mu_r \tau_\mathrm{MW}} A \tilde{P}_\parallel^\sigma \cos\tilde{\omega} \tilde{t}.
\end{equation}
Equations \eqref{eq:P-AC1z0} and \eqref{eq:P-AC1x0} are then rewritten as follows:
\begin{align}
    \frac{\dif {\tilde{P}_z^\sigma}}{\dif \tilde{t}}+\tilde{P}_z^\sigma &= -4\pi \left\{(\chi^\infty + 1/2)\cos\tilde{\omega} \tilde{t} - \chi^\infty \tilde{\omega}\sin\tilde{\omega} \tilde{t} \right\} \notag \\
    & \qquad + A (\tilde{P}_\parallel^\sigma)^2 \cos\tilde{\omega} \tilde{t},  \label{eq:P-AC1z} \\
    \frac{\dif {{\bf \tilde{P}}_\parallel^\sigma}}{\dif \tilde{t}}+{\bf \tilde{P}}_\parallel^\sigma &= - A \tilde{P}_z^\sigma {\bf \tilde{P}}_\parallel^\sigma \cos\tilde{\omega} \tilde{t}.  \label{eq:P-AC1x}
\end{align}

We numerically solved Eqs.~\eqref{eq:P-AC1z} and \eqref{eq:P-AC1x} with $\frac{\omega}{2\pi} = 300\unit{Hz}$ and $\tau_{\mathrm{MW}}=1\unit{ms}$. Note that the direction of ${\bf \tilde{P}}_\parallel^\sigma(t)$ in Eq.~\eqref{eq:P-AC1x} is determined by the spontaneous symmetry breaking and changes by the effective rotational diffusion.
Figure~\ref{fig:singleth}a displays the time evolution of $\tilde{P}_{\parallel}^{\sigma}(t), \tilde{P}_z^{\sigma}(t)$, and $v(t)$, from the initial condition $(\tilde{P}_{\parallel}^{\sigma}(0),\, \tilde{P}_z^{\sigma}(0))=(0.1,0.1)$.
The polarization $(\tilde{P}_{\parallel}^{\sigma}, \tilde{P}_z^{\sigma})$ reaches the periodic state, in which $\tilde{P}_z^{\sigma}(t)$ oscillates with the period identical to that of the external field, but $\tilde{P}_{\parallel}^{\sigma}(t)$ does so with the half period (fundamental frequency $2\omega$) without changing its sign.
As a result, the velocity $v(t)$ oscillates at the period of the external field, corresponding to the reciprocating motion as observed experimentally.
This rolling mechanism of an AC Quincke roller is depicted in Fig.~\ref{fig:singleth}b.
Physically, when the external field is applied, as in the DC case, the component $\tilde{P}_{\parallel}^{\sigma}$ is initially developed by instability due to ionic flow around the particle and reaches a characteristic strength.
In the AC case, the field is reversed every half period and the ionic flow is also reversed accordingly in the $z$-direction, but the charge distribution remains biased in such a way that the direction (sign) of $\tilde{P}_{\parallel}^{\sigma}$ is conserved.
Figure~\ref{fig:singleth}c shows the trajectory in $(\tilde{P}_{\parallel}^{\sigma}, \tilde{P}_z^{\sigma})$ space, which goes to the red closed loop (limit cycle) after a few cycles.
Trajectories from different initial conditions went to the same limit cycle as long as the sign of $\tilde{P}_{\parallel}^{\sigma}(0)$ was positive.
If $\tilde{P}_{\parallel}^{\sigma}(0)<0$, the limit cycle to realize was the one obtained as a mirror image (with respect to the $\tilde{P}_z$ axis) of the limit cycle shown in \figref{fig:singleth}c.
This is because, if $(\tilde{P}_{\parallel}^{\sigma}(t), \tilde{P}_z^\sigma(t))$ is a solution to Eqs.~\eqref{eq:P-AC1z} and \eqref{eq:P-AC1x}, $(-\tilde{P}_{\parallel}^{\sigma}(t), \tilde{P}_z^\sigma(t))$ is also a solution.
In our model, this relaxation to the limit cycle and the resulting reciprocating motion were observed in a broad range of the frequency, from the low frequency limit $f\rightarrow 0$ up to $f\sim 1/\tau_{\mathrm{MW}}$.
For higher frequencies, the limit cycle does not result in reciprocating motion because $\tilde{P}_{\parallel}^{\sigma}(t)=0$ (see Fig. S1, ESI\dag).

To compare the theory and the experimental result further, the autospectral densities of $\tilde{P}_{\parallel}^\sigma(t)$, $\tilde{P}_z(t)$ and $v(t)$ in the periodic state of the model are shown in Fig.~\ref{fig:singleth}d.
In the model, $\tilde{P}_{\parallel}^{\sigma}(t)$ and $\tilde{P}_z^\sigma(t)$ have only even and odd harmonics respectively.
Then the velocity $v(t)$ has only odd harmonics according to Eq.~\eqref{eq:v_AC}, as confirmed in Fig.~\ref{fig:singleth}d.
Since the third harmonic peak of $S_{vv}$ is about $1/100$ of the peak at the fundamental frequency $300\unit{Hz}$ (Fig.~\ref{fig:singleth}d), it is reasonable that we did not see the odd higher harmonics in the experiment (Fig.~\ref{fig:single}b).
In contrast, unlike the experiment, $S_{vv}(f)$ in the model does not show a low-frequency peak, implying that the reciprocating motion in this model does not involve the long-time persistent motion that was observed experimentally (Sec.~\ref{sec:singleE}).
We consider that this is because the model does not capture the asymmetry of the surface charging rate at the upside and downside of the particle near the electrode correctly. Zhang {\it et al.} pointed out this asymmetry and took it into account in their model, which successfully explained the spontaneous oscillation of the Quincke roller under a strong DC field \cite{Zhang2021}.

Here we introduce a biased model that incorporates the top-bottom asymmetry.
For simplicity, we consider the effect of the asymmetry by adding a constant biased term to the polarization in $z$-direction:
$P_z(t)=P_z^{\sigma}(t)+P_z^{\epsilon}(t)+P_z^{\alpha}$, where $P_z^{\alpha}=4\pi\epsilon_0 E_0 \alpha$ with a constant dimensionless value $\alpha$.
This modifies Eqs.~\eqref{eq:P-AC1z} and \eqref{eq:P-AC1x} as follows:
\begin{align}
    \frac{\dif {\tilde{P}_z^\sigma}}{\dif \tilde{t}}+\tilde{P}_z^\sigma &= -4\pi \left\{(\chi^\infty + 1/2)\cos\tilde{\omega} \tilde{t} + \chi^\infty \alpha - \chi^\infty \tilde{\omega}\sin\tilde{\omega} \tilde{t} \right\} \notag \\
    & \qquad + A (\tilde{P}_\parallel^\sigma)^2 \cos\tilde{\omega} \tilde{t},  \label{eq:P-AC1z_b} \\
    \frac{\dif {{\bf \tilde{P}}_\parallel^\sigma}}{\dif \tilde{t}}+{\bf \tilde{P}}_\parallel^\sigma &= - A {\bf \tilde{P}}_\parallel^\sigma \cos\tilde{\omega} \tilde{t} (\tilde{P}_z^\sigma + 4 \pi \chi^\infty \alpha).  \label{eq:P-AC1x_b}
\end{align}
Note that these equations are not equivalent to the case with a nonzero DC electric field ${\bf E}(t) = {\bf E}_0 \cos\omega t + {\bf E}_1$ even formally. There is no DC component in the r.h.s. of Eqs.~\eqref{eq:P-AC1z_b} and \eqref{eq:P-AC1x_b}, but the DC component appears due to nonlinearity of the equations. Another note is that the steady solution of these equations in the DC limit ($\omega\rightarrow 0$) just results in the shift of the stable fix point from that of Eqs.~\eqref{eq:P-AC1z0} and \eqref{eq:P-AC1x0}.
The modified equations (Eqs.~\eqref{eq:P-AC1z_b} and \eqref{eq:P-AC1x_b}) indicate that both $\tilde{P}_\parallel^\sigma$ and $\tilde{P}_z^\sigma$ can be neither odd nor even functions. It means that $\tilde{P}_\parallel^\sigma$ has a nonzero odd component which generates even components in the velocity. The autospectra of the biased model with $\alpha=1$ are shown in Fig.~\ref{fig:singleth}e. The polarization and velocity have both even and odd peaks. Figure ~\ref{fig:singleth}f shows the positions of a particle $x(t)$ for the models with no bias ($\alpha=0$) showing no net motion and a bias ($\alpha=0$) showing a net motion. Because there is a nonzero DC component of the velocity in the biased model, the reciprocating motion accompanies net motion.
Although this model does not involve the rotational diffusion, the existence of DC component of velocity leads to the Lorentzian by considering the athermal rotational coefficient which we discussed in Sec.~\ref{sec:singleE}. Therefore, the ABP-like long-time persistent motion we observed experimentally (Sec.~\ref{sec:singleE}) is accounted for by the asymmetry due to the electrode.
% Actually, we observed the similar the low-frequency modes under different external frequencies ($200\unit{Hz}$, $250\unit{Hz}$, and $300\unit{Hz}$; see Fig.~S2, ESI\dag), though they can depend on the external frequencies.
% We therefore consider that the coupling to the spontaneous oscillation reported by Zhang {\it et al}.\ may lead to the long-time persistent motion we observed.

\section{Multi-particle behavior}  \label{sec:MultiParticle}

\begin{figure*}[t!]
\centering
\includegraphics[width=14.1cm]{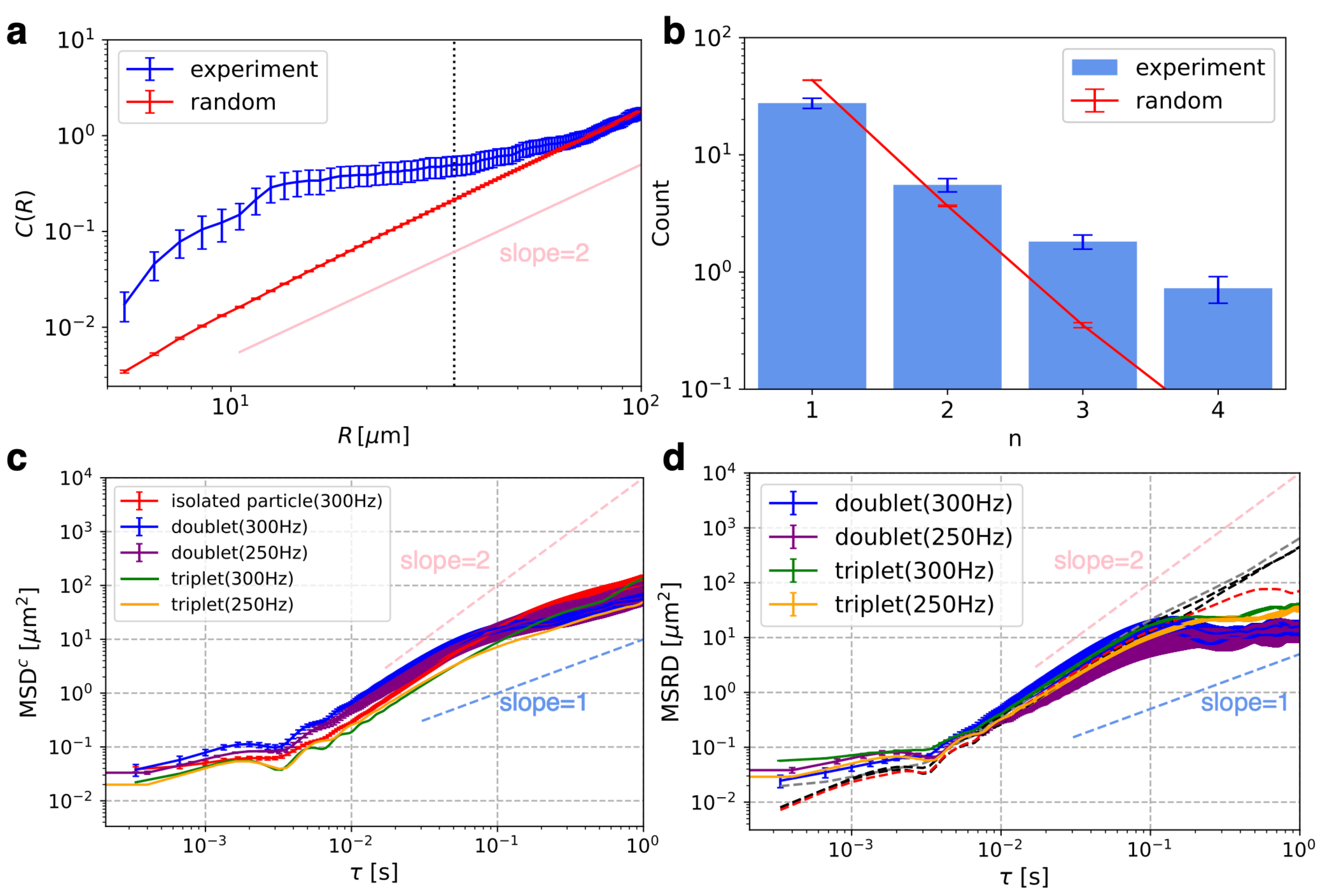}
\caption{Dynamic clustering and bound states of doublets and triplets. (a)	Correlation integral $C(R)$ for the experiment ($250\unit{Hz}$, blue) and for the numerically generated, randomly distributed particles (red). We used low-magnification (x10 objective lens) time-lapse images, taken at an interval of $100T$ to ensure statistical independence between images. The time-averaged number of particles was $\overline{N}_p \simeq 52$. Error bars are the standard errors from $10$ slices. In simulations, the same number of particles were randomly placed in a box of the same size ($1031.2\unit{\mu m}$ square), and average is taken over $10000$ realizations. Error bars are the standard errors from those realizations. The vertical dotted line indicates $R=r_\mathrm{th}=35\unit{\mu m}$.
(b)	Cluster size distribution for the experiment ($250\unit{Hz}$, blue) and for the numerically generated random particles (red). Same data as (a) are used both for the experiment and for the random case.
(c)	The MSDs of isolated particles, and the MSDs of the centroids of the particles forming doublets and triplets: $\mathrm{MSD}^{c}(\tau)$.
The averages over a few samples are shown.
The number of samples was three for isolated particles, two for doublets ($300\unit{Hz}$), one for triplet ($300\unit{Hz}$), three for doublets ($250\unit{Hz}$), and one for triplet ($250\unit{Hz}$).
(d)	The MSRD of the particle pairs forming doublets and triplets (solid line): $\mathrm{MSRD}(\tau)$. For the triplets, averages of $\mathrm{MSRD}(\tau)$ over all pairs $i,j=1,2,3$ are shown. Sample average was also taken, as in (c). The dashed gray line is $\mathrm{MSRD}(\tau)$ for a particle pair that initially formed a doublet, but collapsed during the observation time. The dashed black lines show $\mathrm{MSRD}(\tau)$ between a single particle and one of the particles forming a doublet, which join during the observation time to form a triplet (see Video~S12, ESI\dag). From the same set of the three particles, $\mathrm{MSRD}(\tau)$ between the two particles that constituted the doublet at the beginning is shown by the dashed red line.
}
\label{fig:clustering}
\end{figure*}

Besides single-particle motion, we also observed formation of clusters of multiple particles, which were then maintained over 100--1000 periods or longer in our dilute system (see Video~S3, ESI\dag). To confirm the cluster formation, we first check spatial distribution of particles in a wider field of view ($1031.2\unit{\mu m}$ square region) using a x10 objective lens (Olympus, UPlanFL10X, NA$0.30$) and at $250\unit{Hz}$.
We measure the radial distribution function
\begin{equation}
g(r)=\frac{1}{N_p(N_p-1)}\sum_{i\neq j} \delta (|{\bf x}_{ij}|-r),
\end{equation}
where $N_p$ is the number of the particles in the field of view, and ${\bf x}_{ij}:={\bf x}_{i}-{\bf x}_{j}$ with ${\bf x}_{i}$ being the position of the $i$th particle.
Then we evaluate the correlation integral $C(R)$ \cite{Grassberger1983},
\begin{equation}
    C(R)=\int_0^{R} g(r) \dif r
\end{equation}
and compare with that of randomly placed particles, generated numerically by the uniform distribution.
As shown in \figref{fig:clustering}a, in the uniformly random case, $C(R)$ increases with a power law $R^D$ with $D=2$ in the two-dimensional case \cite{Grassberger1983}. By contrast, the experimental result differs significantly. The number of pairs with short interparticle distance is significantly larger, indicating the formation of clusters. Based on this result, we consider that a pair of particles forms a cluster if the interparticle distance is shorter than $r_\mathrm{th}=35\unit{\mu m}$ (black dotted line in \figref{fig:clustering}a, chosen in such a way that $C(R)$ for $R<r_\mathrm{th}$ is markedly different from that of the uniformly random case).
Using this threshold, we obtain the cluster size distribution (the number of clusters of size $n$, evaluated by the freud library \cite{freud2020}).
The result is shown in \figref{fig:clustering}b, with the comparison to the uniformly random case. This shows that clusters ($n \geq 2$) are indeed formed more frequently than the uncorrelated random case.

Now we focus on clusters with $n=2$ (doublets; see Videos~S4--S7, ESI\dag) and $n=3$ (triplets; see Videos~S8--S11, ESI\dag).
First, note that such clusters are dynamic: clusters are not trapped at fixed locations on the substrate, but instead each constituent particle moves similarly to isolated particles. This is confirmed by the MSD of the centroid of the particles forming a doublet or a triplet:
\begin{equation}
     \mathrm{MSD}^{c}(\tau)=\frac{1}{N-m}\sum_{k=1}^{N-m}({\bf x}^{c}(t_{k+m})-{\bf x}^{c}(t_{k}))^2,
\end{equation}
where ${\bf x}^{c}:=({\bf x}_1+{\bf x}_2)/2$ for doublets and ${\bf x}^{c}:=({\bf x}_1+{\bf x}_2+{\bf x}_3)/3$ for triplets, and the lag time is $\tau = (m/3000)\unit{s}$ for $300\unit{Hz}$ and $\tau = (m/2500)\unit{s}$ for $250\unit{Hz}$.
This MSD shows oscillations in the short-time region and superdiffusion in a relatively long time region, similarly to that of isolated particles (\figref{fig:clustering}c).

Doublets and triplets are bound for some length of times. To confirm the binding, we define the mean squared relative distance (MSRD) of the constituent particles forming a doublet or a triplet:
\begin{equation}
     \mathrm{MSRD}(\tau)=\frac{1}{N-m}\sum_{k=1}^{N-m}(|{\bf x}_{ij}(t_{k+m})|-|{\bf x}_{ij}(t_{k})|)^2,
\end{equation}
which quantifies if the relative distance grows or not.
The values of $\mathrm{MSRD}(\tau)$ for such doublets and triplets that kept $r<r_\mathrm{th}$ during the observation time ($1091T$) are shown in \figref{fig:clustering}d (solid line). All show the sign of saturation, suggesting that the particles are bound, more visibly for doublets than triplets.
However, the binding state of doublets and triplets does not last indefinitely; they can also collapse spontaneously or by an approach of other particles. The dashed gray line in \figref{fig:clustering}d shows $\mathrm{MSRD}(\tau)$ of the doublet that collapsed during the same observation time. The dashed black lines show $\mathrm{MSRD}(\tau)$ between a single particle and one of the particles forming a doublet, which join during the observation time to form a triplet (see Video~S12, ESI\dag).
These datasets of $\mathrm{MSRD}(\tau)$ do not saturate because the corresponding interparticle distance did not always satisfy $r<r_\mathrm{th}$ during the observation time.
In contrast, from the same triplet that was formed by merging of a doublet and a single particle, $\mathrm{MSRD}(\tau)$ for the two constituent particles of this doublet saturated (dashed red line in \figref{fig:clustering}d) as expected, because this pair was bound during the entire observation time.
Note that only two out of eleven clusters (doublets and triplets) showed spontaneous collapse during our observation time ($1091T$); this suggests that their life time is in the order of $10^3T$ or so.

In the following, we characterize structure and dynamics of a doublet and a triplet (Secs.~\ref{sec:doublet} and \ref{sec:triplet}.), compare the motion of a constituent particle with that of an isolated particle (Sec.~\ref{sec:comparison}), and discuss the interaction between constituent particles (Sec.~\ref{sec:Discussion_M}).

\subsection{Doublet}  \label{sec:doublet}

\begin{figure*}[t!]
\centering
  \includegraphics[width=18.1cm]{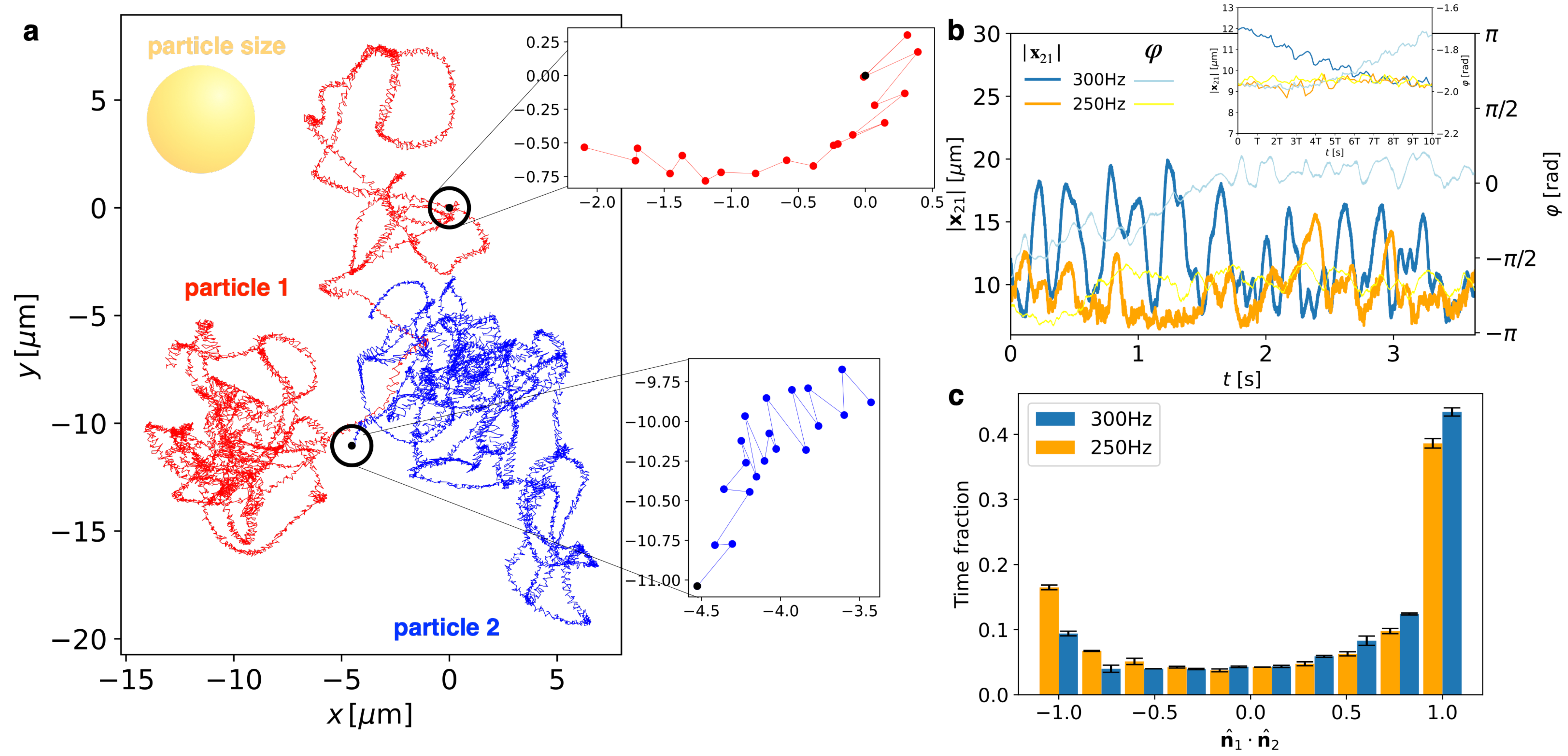}
  \caption{Experimental results of doublets ($150\unit{V}$). (a) An example of trajectories of a doublet ($300\unit{Hz}$, over $1091$~periods $= 3.639\unit{s}$). The insets show the positions recorded every half period for the first $10$ periods. (b) Distance $|{\bf x}_{21}(t)|$ and relative angle $\varphi(t)$ (azimuth of ${\bf x}_{21}(t)$) between the two particles of the doublet shown in (a).  The inset shows an enlargement displaying the first $10$ periods, with different magnification rates for $|{\bf x}_{21}(t)|$ and $\varphi(t)$. (c) Histograms of the inner product of the self-propelling directions of doublets, $\hat{\bf n}_1 \cdot \hat{\bf n}_2$. Errorbars are the standard errors evaluated from two doublets ($300\unit{Hz}$) and three doublets ($250\unit{Hz}$).}
\label{fig:doublet}
\end{figure*}

First we focus on doublets, which were at least $100\unit{\mu m}$ away from other particles.
We analyzed five doublets (two for $300\unit{Hz}$ and the other three for $250\unit{Hz}$) which were kept bound during the whole observation time $1091T$. Trajectories of the two particles of a doublet ($300\unit{Hz}$) are shown in Fig.~\ref{fig:doublet}a, as an example.
The insets show the positions recorded every half period for the first 10 periods.
Similarly to isolated particles, the doublet constituents also exhibited short-time reciprocating behavior (at the frequency of the external field) and long-time persistent motion (see Videos~S4--S7, ESI\dag).
Despite the persistent motion of each particle, the interparticle distance $|{\bf x}_{21}|$ was maintained in some range as shown in Fig.~\ref{fig:doublet}b, with noticeable oscillations at a few Hz
in addition to the smaller ones at the external frequency (Fig.~\ref{fig:doublet}b inset).
The relative angle $\varphi(t)$, defined by the azimuth of ${\bf x}_{21}$, also changed in time (green curve and right axis of Fig.~\ref{fig:doublet}b).
To characterize the correlation between the motions of the two particles, we define the unit vector indicating the self-propelling direction over a half period, called the propulsion vector hereafter, $\hat{\bf n}_i(t_m) := \Delta {\bf x}_i(t_m) / |\Delta {\bf x}_i(t_m)|$ with $\Delta {\bf x}_i(t_m) := {\bf x}_i(t_{m+1})-{\bf x}_i(t_m)$. ($t_m:=mT/2$, $m=0,1,2\dots$).
Figure~\ref{fig:doublet}c shows the histograms of the inner product $\hat{\bf n}_1\cdot\hat{\bf n}_2$, indicating that the two particles in a doublet tend to align their self-propelling directions, mostly parallel but sometimes anti-parallel.
This suggests that they also tend to align their polarization vectors ${\bf P}_\parallel^\sigma$, according to Eq.~\eqref{eq:v_AC} of the single-particle theory.
Further discussions on the interaction will be given in Sec.~\ref{sec:Discussion_M}.

\subsection{Triplet}  \label{sec:triplet}

\begin{figure*}[t!]
\centering
  \includegraphics[width=18.1cm]{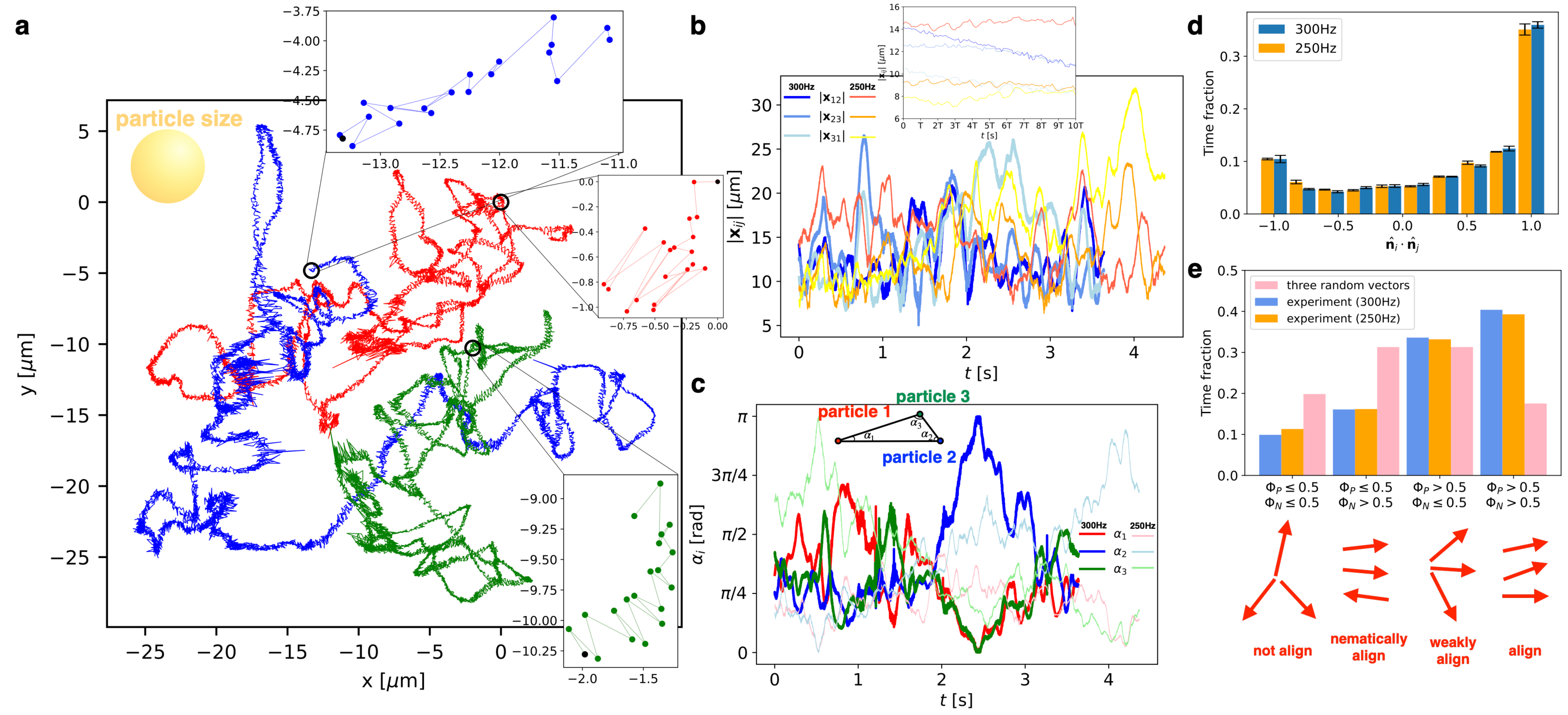}
  \caption{Experimental results of triplets ($150\unit{V}$). (a) An example of trajectories of the three particles forming a triplet (over $1091$~periods $= 3.639\unit{s}$, $300\unit{Hz}$). The insets show the positions recorded every half period for the first $10$ periods. (b) Distance between each pair, $|{\bf x}_{ij}(t)|$. The inset shows an enlargement displaying the first $10$ periods. ($300\unit{Hz}$ and $250\unit{Hz}$) (c) Inner angles of the triplet, $\alpha_i(t)$. ($300\unit{Hz}$ and $250\unit{Hz}$) (d) Histogram of the inner products of the propulsion vectors. Errorbars are taken for three constituents of a triplet. (e) Histograms of the polar and nematic order parameters of the propulsion vectors, in the experiment (blue: $300\unit{Hz}$, orange: $250\unit{Hz}$) and in the case of three random unit vectors (pink). }
\label{fig:triplet}
\end{figure*}

\begin{figure}[t!]
\centering
  \includegraphics[width=8.1cm]{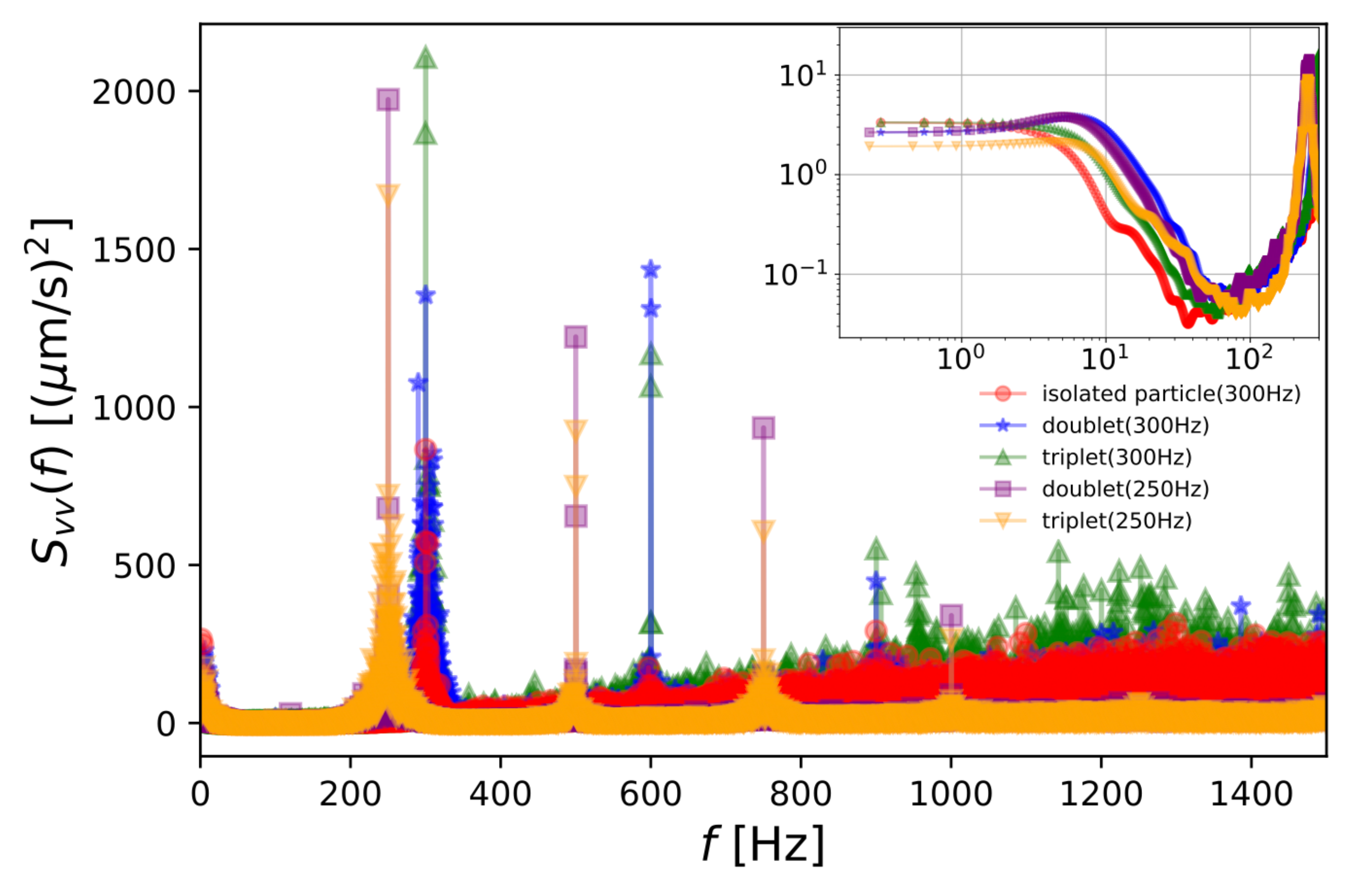}
  \caption{Comparison of energy spectral density among isolated single particles (red), particles forming doublets (blue: $300\unit{Hz}$, purple: $250\unit{Hz}$), and particles forming triplets (green: $300\unit{Hz}$, orange: $250\unit{Hz}$). For each spectrum, average is taken over all directions ($v_x$ and $v_y$) and all constituent particles of all samples. Three isolated particles, two doublets at $300\unit{Hz}$, three doublets at $250\unit{Hz}$, one triplet at $300\unit{Hz}$ and one triplet at $250\unit{Hz}$ are used. The inset shows an enlargement of a low-frequency region. To reduce statistical fluctuations, the time series were divided into $30$ segments (each segment contains $36$ periods), then the power spectra were averaged.}
  \label{fig:comparison}
\end{figure}

Here we focus on triplets, which were at least $100\unit{\mu m}$ away from other particles.
We analyzed two triplets (one for $300\unit{Hz}$ and the other for $250\unit{Hz}$) which were kept bound during the whole observation time $1091T$.
Particle trajectories are shown in Fig.~\ref{fig:triplet}a.
The insets show the positions recorded every half period for the first 10 periods.
Similarly to isolated particles and doublets, the particles forming a triplet also showed short-time reciprocating behavior and long-time persistent motion (see Videos~S8--S11, ESI\dag).
The interparticle distance $|{\bf x}_{ij}|$ of any pair $i,j = 1,2,3$ was maintained in some range as shown in Fig.~\ref{fig:triplet}b.
%, with small oscillations in the period (See the inset).
The geometric configuration was also characterized in terms of the inner angles $\alpha_{i}$ of the triangle, whose vertices are set by the positions of the three particles (see the sketch in Fig.~\ref{fig:triplet}c).
The result in Fig.~\ref{fig:triplet}c shows that the triangle tends to be obtuse most of the time.
The propulsion vectors $\hat{\bf n}_i$ are also defined as in the doublet case and analyzed.
Figure~\ref{fig:triplet}d shows the histogram of the inner products $\hat{\bf n}_i \cdot \hat{\bf n}_j$.
This indicates that any pair of particles tends to align their self-propelling directions, mostly parallel but sometimes anti-parallel.
To further characterize the relation among three propulsion vectors, we use the following two order parameters: the polar order parameter $\Phi_P$ and the nematic order parameter $\Phi_N$, defined by
\begin{eqnarray}
  \Phi_P=\left|\frac{1}{3}\sum_{k=1}^3e^{i\theta_k}\right|, \qquad \Phi_N=\left|\frac{1}{3}\sum_{k=1}^3e^{2i\theta_k}\right|, \label{eq:order}
\end{eqnarray}
where $\theta_k$ is the azimuth of the propulsion vector $\hat{\bf n}_k$.
We consider four states as shown in Fig.~\ref{fig:triplet}e, characterized by $\Phi_P$ and $\Phi_N$ being $\leq 0.5$ or $> 0.5$, and measured the fraction of time spent in each state for $300\unit{Hz}$ and $250\unit{Hz}$ (Fig.~\ref{fig:triplet}e).
For comparison, the time fraction for the case of three randomly generated unit vectors is also shown.
The result in Fig.~\ref{fig:triplet}e shows that the triplet tends to be in the aligned state ($\Phi_P, \Phi_N > 0.5$).

\subsection{Comparison between isolated and clustering particles}  \label{sec:comparison}

As we have seen, particles show similar short-time reciprocating motion and long-time persistent motion, regardless of whether they form a cluster or not.
We discuss similarities and dissimilarities between isolated particles and clustering particles here. First, let us see the MSDs for isolated particles as well as the centroids of the doublets and triplets as already shown in \figref{fig:clustering}c.
While all these results show oscillations in the short-time region and superdiffusion in a relatively long time region, we also notice a few differences.
First, the oscillations in the short-time region are more prominent for the triplet, then for the doublet, and least prominent for the isolated particles.
This suggests that the particle interaction contributes to the larger amplitude of the reciprocating motion.
Moreover, the time region showing the superdiffusive exponent is shorter for the clusters than for the isolated particles.
This is presumably because of a binding effect of the clusters.

We also compare the energy spectral density (Fig.~\ref{fig:comparison}).
Similarly to that of the isolated particles, the spectra for the doublets and triplets also show a sharp peak at the external frequency as well as a low-frequency Lorentzian mode (inset of Fig.~\ref{fig:comparison}).
Besides, however, the spectra for the doublet and triplet also show the considerable second and third harmonic peaks ($600\unit{Hz}$ and $900\unit{Hz}$ for $300\unit{Hz}$, $500\unit{Hz}$ and $750\unit{Hz}$ for $250\unit{Hz}$, respectively) which were absent or invisible in the isolated case.
Since the third harmonic peak is theoretically expected to appear even in the isolated case (see Sec.~\ref{sec:singleT} and \figref{fig:singleth}d), we consider that it became visible here because of the reduction of the background noise in the high frequency region (see \figref{fig:comparison}).
This reduction of noise is possibly because the interaction increased the stability of the dynamics of the polarization and the velocity.
In contrast, the apparent second harmonic peak is characteristic of the clustering particles; for isolated particles, it did not appear in the experiment, and is expected to be very weak in theory even if we consider the biased single-particle model in Sec.~\ref{sec:singleT}. More precisely, the strength of the second harmonic peak in this model is no more than $1/1000$ of that of the fundamental frequency, so that it is reasonable that we did not see it. Therefore, the considerable second harmonic peak is inferred to be due to the interaction discussed in the next subsection.

\subsection{Discussions on the interaction between particles}\label{sec:Discussion_M}

Similarly to the DC case, Quincke rollers under an AC field also interact through hydrodynamic flow and electrostatic effect.
As described in Bricard \textit{et al.\ } \cite{Bricard2013},
the hydrodynamic flow generated by a roller at ${\bf x}_j$ is a dipolar flow for $x_{ij} := |{\bf x}_{ij}| \gg H$, where $H$ is the cell thickness ($H=30\unit{\mu m}$ here).
In the vicinity of the particle ($x_{ij} \lesssim H$), aligning flow is produced \cite{Bricard2013}.
The reciprocating motion under the AC field simply makes the flow field oscillate.
On the electrostatic interaction, the parallel component of the electric field at ${\bf x}_i$ induced by the polarization ${\bf P}^{(j)}$ of particle $j$, denoted by $\delta {\bf E}_{\parallel}^{(j)}({\bf x}_i,t)$, is given by \cite{Bricard2013}
\begin{equation}\label{eq:electrostatic_interaction}
  \delta{\bf E}_{\parallel}^{(j)}({\bf x}_i,t)=\frac{3}{2\pi\epsilon_0 x_{ij}^3}\left[\frac{a}{x_{ij}}P_z^{(j)} \hat{{\bf x}}_{ij}-\frac{a^2}{x_{ij}^2}{\bf P}_{\parallel}^{(j)}\cdot(5\hat{{\bf x}}_{ij}\hat{{\bf x}}_{ij}-{\bf I})+\mathcal{O}(\frac{a^3}{x_{ij}^3}) \right],
\end{equation}
with $\hat{{\bf x}}_{ij} := {\bf x}_{ij} / |{\bf x}_{ij}|$ and the identity matrix ${\bf I}$.
Thus, particle $i$ is exposed to the total electric field ${\bf E}_0 \cos\omega t + \sum_j \delta {\bf E}_{\parallel}^{(j)}({\bf x}_i,t)$, which should replace ${\bf E}_0$ in Eqs.~\eqref{eq:eom} and \eqref{eq:electrodynamics}.
Since $P_z^{(j)}$ and ${\bf P}_{\parallel}^{(j)}$ have odd and even harmonics, respectively (see Sec.~\ref{sec:singleT} and \figref{fig:singleth}d), $\delta{\bf E}_{\parallel}^{(j)}({\bf x}_i,t)$ has both of them, and so does ${\bf P}^{\sigma}$ of particle $i$.
This accounts for the considerable second harmonic peak observed in the energy spectral density only for clustering particles (\figref{fig:comparison}).

We consider that the stable formation of clusters is also a result of such hydrodynamic and electrostatic interactions under the AC external field.
Interestingly, clustering has also been reported for DC Quincke rollers with periodic on-off switching \cite{Karani2019} but not under a constant DC field \cite{Bricard2013,Bricard2015}.
Therefore, time-dependent propulsion and interaction may be a key to clustering.
Elucidating the clustering mechanism is an important open problem left for future studies.

\section{Conclusion and outlook}

Our experimental investigation into Quincke rollers under an AC electric field revealed the reciprocating motion at the frequency of the external field, accompanied by ABP-like long-time persistent motion.
The single-particle theory explained how this reciprocating motion arises as a result of oscillations of the polarization, which turned out to have different parities between the vertical and parallel components.
The ABP-like long-time persistent motion is considered to arise by the effect of charging asymmetry at the upside and downside of the particle near the electrode. The extreme deviation between the thermal and observed values of the rotational diffusion coefficient can be attributed to athermal noise caused by the interplay between surface roughness of the electrodes and electrokinetic effect at the surface of particles. We also found formation of clusters such as doublets and triplets, and characterized their structure and dynamics.
We showed that clustering results in the generation of the considerable second harmonic peak in the energy spectral density, which we accounted for by the dipole-dipole interaction between particles.

There are a number of interesting directions to take for future studies of AC Quincke rollers.
Experimentally, it is tempting to study formation of larger clusters in denser suspension, as well as dynamical processes of cluster formation.
Studying the effect of AOT concentration, as reported by Zhang {\it et al.\ } \cite{Zhang2021} for strong DC fields, may also be interesting in the AC case.
The effect of AOT micelles may be incorporated in our model, in terms of position-dependent conductivity as modeled by Zhang {\it et al.\ } \cite{Zhang2021}.
Simulations of a collection of particles interacting through the hydrodynamic and electrostatic interactions formulated by Bricard {\it et al.\ } \cite{Bricard2013}, but under an AC field, are also desired.

From broader perspectives, studies of artificial bipolar SPPs have not been as developed as those for the unidirectional, polar counterparts.
The system of AC Quincke rollers presented in this work may play a unique role in this context, thanks to their regular reciprocation that takes place at the externally controlled frequency. It is known that for some types of self-propelled Janus particles the self-propulsion velocity may be reversed by changing parameters on the external energy source (such as the frequency of the applied voltage, the strength of the light irradiation, etc.) \cite{Vutukuri2020,GomezSolano2017,Mano.etal-PNAS2017,Suzuki2011}, but the propulsion mechanism and the particle interaction are typically very different in the reversed state, so that symmetric reciprocal motion cannot be realized by simply switching between the normal and reversed states back and forth. By contrast, the AC Quincke rollers studied in the present work have advantages that the reciprocal motion is almost symmetric, owing simply to the electric field reversal.
While having fundamental aspects in common with other active matter systems, such as self-propulsion of particles and energy dissipation at multiple scales, periodically driven active systems may serve as a new class of active matter worth exploring.
We hope that it will develop and contribute to deeper understanding of active matter and related phenomena in biology and non-equilibrium physics, and that controllable bipolar SPPs such as AC Quincke rollers will serve as a useful tool in this direction.

\section*{Author Contributions}
%We strongly encourage authors to include author contributions and recommend using \href{https://casrai.org/credit/}{CRediT} for standardised contribution descriptions. Please refer to our general \href{https://www.rsc.org/journals-books-databases/journal-authors-reviewers/author-responsibilities/}{author guidelines} for more information about authorship.
ANK and MS designed the experiment. KAT contributed to continue the research. ANK performed experiments and numerical simulations. ANK and MS contributed to theoretical calculations. All authors discussed and contributed to the analysis and the interpretation of the results. All authors wrote the manuscript.

\section*{Conflicts of interest}
There are no conflicts to declare.%In accordance with our policy on \href{https://www.rsc.org/journals-books-databases/journal-authors-reviewers/author-responsibilities/#code-of-conduct}{Conflicts of interest} please ensure that a conflicts of interest statement is included in your manuscript here.  Please note that this statement is required for all submitted manuscripts.  If no conflicts exist, please state that ``There are no conflicts to declare''.

\section*{Acknowledgements}
We thank D. Nishiguchi for his help to prepare the cell, and H. Wada, N. Yoshinaga, F. Peruani, H. Kitahata for fruitful discussions. We thank Ge-Wei Chen and the analysis center of SJTU for helping to measure surface roughness of ITO film. This work is supported by Grand-in-Aid for JSPS Fellows (Grant No. JP17J06659) and KAKENHI (Grant Nos. JP25103004, JP19H05800, JP20H01826).

%%%END OF MAIN TEXT%%%


%\balance


%%%REFERENCES%%%
%\bibliography{rsc.bib} %You need to replace "rsc" on this line with the name of your .bib file
\bibliographystyle{apsrev4-1}

%\bibliographystyle{rsc} %the RSC's .bst file
%\input{Quincke.bbl}
\end{document}

\maketitle

% \begin{abstract}
% Your abstract.
% \end{abstract}

\section*{Supplementary Note 1: Model of a single self-propelled particle with a sinusoidal and constant velocity}
Consider a particle self-propelling at speed $v(t)$ under the effect of translational and rotational diffusion. The translational diffusion and rotational diffusion coefficients are denoted by $D$ and $D_\theta$, respectively.  Let ${\bf x}$ and $\theta$ be the position and the propelling direction, respectively.  The equations of motion read:
\begin{equation}
\dot{{\bf x}}(t)=v(t)\hat{{\bf n}}(t)+\pmb{\zeta}(t)
\end{equation}
\begin{equation}
\dot{\theta}(t)=\xi(t).
\end{equation}
Here, $\hat{{\bf n}}(t)=(\cos\theta(t),\sin\theta(t))$ is the propelling direction vector. The noises are white Gaussian: $\langle \zeta_i(t)\zeta_j(t')\rangle =4D \delta (t-t')\delta_{ij}$, $\langle \zeta_i(t)\rangle=0$, $\langle \xi(t)\xi(t')\rangle =2D_{\theta}\delta (t-t')$, and $\langle \xi(t)\rangle =0$ for arbitrary $t,t'$.
Here let us assume the self-propulsion velocity $v(t)$ to be
\begin{equation}
  v(t)=v_0+v_1 \cos\omega t.
\end{equation}
We calculate the energy power spectrum of this model via the velocity autocorrelation.
The velocity autocorrelation is calculated as follows.
\begin{eqnarray}
\langle {\bf v}(t_1)\cdot {\bf v}(t_2) \rangle&=& v(t_1)v(t_2)\langle \hat{{\bf n}}(t_1)\cdot \hat{{\bf n}}(t_2)\rangle+4D\delta(t_1-t_2)\nonumber\\
&=&v(t_1)v(t_2)\langle \cos\theta(t_1)\cos\theta(t_2)+\sin\theta(t_1)\sin\theta(t_2)\rangle +4D\delta(t_1-t_2)\nonumber\\
&=&v(t_1)v(t_2)\langle \cos(\theta(t_1)-\theta(t_2))\rangle +4D\delta(t_1-t_2)\nonumber\\
&=&v(t_1)v(t_2)\langle \cos(\Delta_{t_1-t_2}\theta)\rangle +4D\delta(t_1-t_2),
\end{eqnarray}
where $\Delta_\tau \theta := \theta(t+\tau) - \theta(t)$ and the time translation invariance of $\theta(t)$ is taken into account. The probability density of $\Delta_\tau\theta$, denoted by $P(\Delta\theta,\tau)$, has the characteristic function that satisfies $\phi(s,t):=\int_{-\infty}^{\infty}\dif \theta P(\theta,t)e^{i\theta s}=e^{-D_{\theta}s^2 t}$.
For $\tau>0$,
\begin{eqnarray}
\langle \cos(\Delta_{\tau}\theta)\rangle &=&\int P(\theta,\tau)\cos\theta \dif \theta \nonumber\\
&=& \frac{\phi(1,\tau)+\phi(-1,\tau)}{2}\nonumber\\
&=& e^{-D_{\theta}\tau}.
\end{eqnarray}
Therefore, neglecting the translation diffusion, we obtain the following expression for the velocity autocorrelation:
\begin{eqnarray}\label{eq:vv}
\langle {\bf v}(t_1)\cdot {\bf v}(t_2) \rangle &=& v(t_1)v(t_2) e^{-D_{\theta}|t_1-t_2|}\nonumber\\
&=& \left[ v_0^2+v_0v_1\left\{\cos(\omega t_1)+\cos(\omega t_2)\right\}+v_1^2\cos(\omega t_1)\cos(\omega t_2) \right] e^{-D_{\theta}|t_1-t_2|}
\end{eqnarray}
which cannot be expressed as a function of $(t_1-t_2)$.  In other words, $\langle {\bf v}(t_1)\cdot {\bf v}(t_2) \rangle$ is not invariant under time translation.

Generally, the energy spectral density is calculated by the Fourier transformation of the velocity autocorrelation $\langle {\bf v}(t)\cdot {\bf v}(t+\tau) \rangle$.  In our model, however, $\langle {\bf v}(t)\cdot {\bf v}(t+\tau) \rangle$ cannot be expressed by a function of $\tau$ only. Therefore, here we consider the periodically-averaged velocity correlation $C(\tau):=\overline{\langle {\bf v}(t)\cdot {\bf v}(t+\tau) \rangle}$, where the overline denotes averaging over a period of the external field. This is evaluated as follows:
\begin{eqnarray}
C(\tau)&=&\left( v_0^2+ v_1^2 \overline{\cos(\omega t)\cos(\omega (t+\tau))}\right) e^{-D_{\theta}\tau}\nonumber\\
&=&\left( v_0^2+ \frac{v_1^2}{2}\overline{(\cos(2\omega t+\omega \tau)+\cos(\omega \tau))}\right) e^{-D_{\theta}\tau}\nonumber\\
&=& \left(v_0^2+\frac{v_1^2}{2}\cos(\omega \tau)\right)e^{-D_{\theta}\tau}.\label{eq:pav_vv}
\end{eqnarray}

Then, the spectral density $S(f)$ is obtained by Fourier transform of Eq.~\eqref{eq:pav_vv}, as follows:
\begin{equation}\label{eq:simple_spectrum}
  S(f)=v_0^2 \frac{2D_{\theta}}{D_{\theta}^2+(2\pi f)^2}+\frac{v_1^2}{2}\left(\frac{D_{\theta}}{D_{\theta}^2+(\omega+2\pi f)^2}+\frac{D_{\theta}}{D_{\theta}^2+(\omega-2\pi f)^2}\right),
\end{equation}
when the translational diffusion is neglected. The translational diffusion adds a constant terms to Eq.~\eqref{eq:simple_spectrum}. For low frequency $f$, the first term is dominant in Eq. ~\eqref{eq:simple_spectrum}, which is Lorentzian with the cutoff frequency $f \sim \frac{D_\theta}{2\pi}$.

To estimate $D_\theta$, we consider the experimentally observed low-frequency region to be a simple Lorentzian: the first term of Eq.~\eqref{eq:simple_spectrum} $v_0^2 \frac{2D_{\theta}}{D_{\theta}^2+(2\pi f)^2}=\frac{2v_0^2}{D_\theta}\cdot\frac{1}{1+(2\pi f/D_\theta)^2}$. The coefficient $\frac{2v_0^2}{D_\theta}$ was determined by the value at $f=0$. By fitting the data for $f<40\unit{Hz}$, we obtained $D_\theta\sim 31.4\pm 0.6~\unit{s}^{-1}$ ($D_\theta^{-1}\sim 0.03~\unit{s}$). Note that this effective rotational diffusive coefficient $D_\theta$ is athermal, can be originated from such as surface heterogeneity of the electrode and particles. (Cf. thermal rotational diffusive coefficient: $D_\theta^{\mathrm{thermal}}=\frac{k_\mathrm{B}T}{8\pi a^3\eta}\sim 10^{-3}\unit{s}^{-1}$ with the room temperature $T$ and the viscosity of hexadecane $\eta$.)
%\vspace{}

\section*{Supplementary figures}

\begin{figure}[h]
\begin{center}
\includegraphics[width=15.1cm]{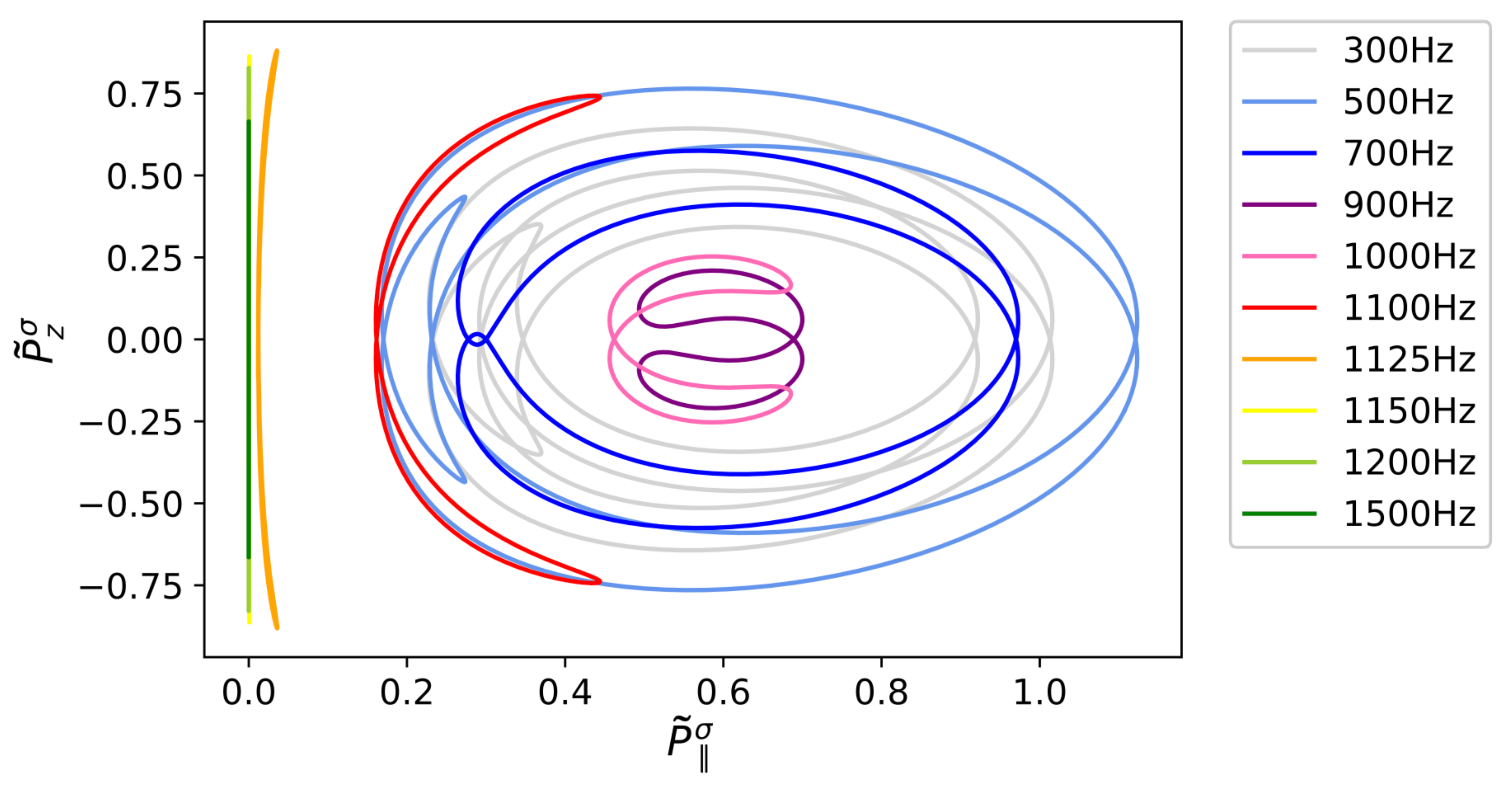}
\caption{Periodic states of the single-particle model for different frequencies of the external electric field. Trajectories from initial conditions with $\tilde{P}_{\parallel}^{\sigma}(0)>0$ are shown, in $(\tilde{P}_{\parallel}^{\sigma}, \tilde{P}_z^{\sigma})$ space. For high enough frequencies, $\tilde{P}_{\parallel}^{\sigma}$ vanishes and the particle does not reciprocate (see Eq.(10) in the main paper).}
\label{fig:sup_LC}
\end{center}
\end{figure}

\newpage

\begin{figure}[h]
\begin{center}
\includegraphics[width=15.1cm]{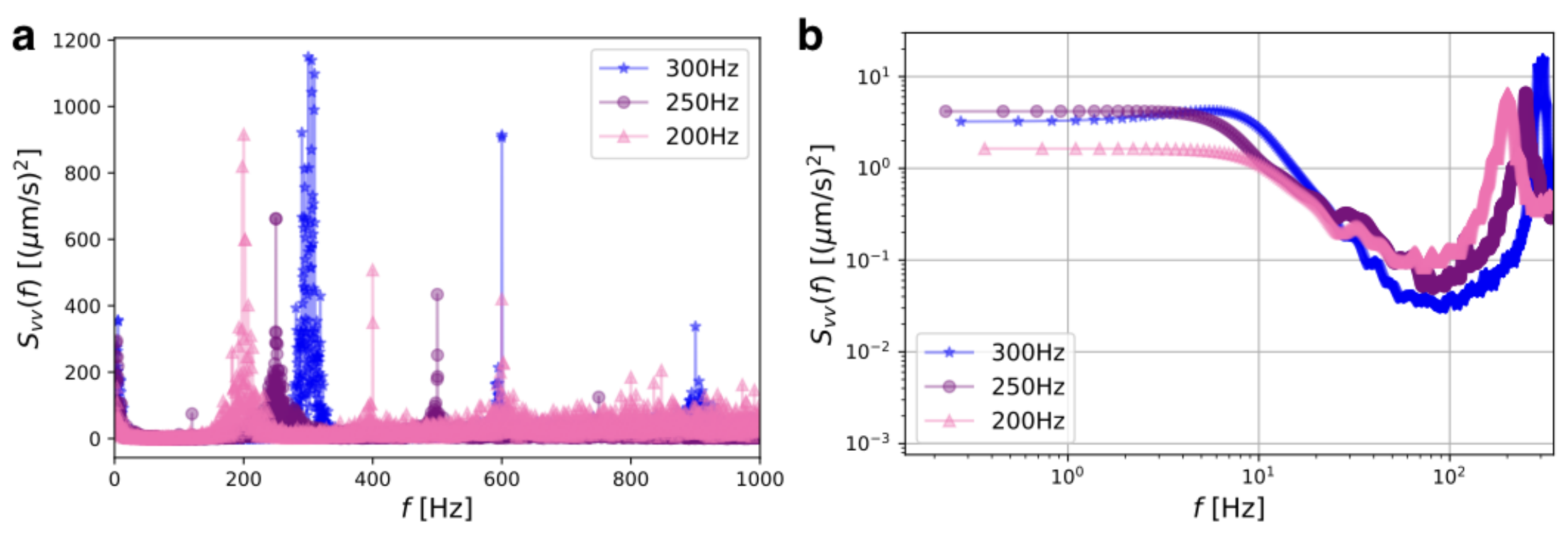}
\caption{Experimental results of the energy spectral density of particles forming a doublet, for different frequencies of the external electric field. The amplitude was kept at $150\unit{V}$ (root-mean-square amplitude). The motion of the particles was captured at the frame rate $10$ times faster than the external frequency (e.g., $2500\unit{fps}$ for $250\unit{Hz}$). The highest peak of each spectrum is found at the frequency of the external field, corresponding to the reciprocating motion of the particles. The second harmonic peak due to the dipole-dipole interaction is also seen in all cases. Panel (b) shows an enlargement of (a) in a low-frequency region. To reduce statistical fluctuations, the time series were divided into $30$ segments (each segment contains $36$ periods), then the power spectra were averaged. All spectra show a similar low-frequency mode. }
\label{fig:sup_otherfreq}
\end{center}
\end{figure}

\section*{Supplementary Note 2: Possible effects of surface roughness of ITO electrode and its interaction with particle surface}
To discuss the possibility that the athermal value of the rotational diffusion coefficient results from surface heterogeneity of the electrode, we measured the surface roughness of the indium-tin oxide (ITO) film on the glass substrate used in the experiments (Mitsuru Optical Co. Ltd.) with an atomic force microscope (AFM, Bruker, Dimension FastScan Bio). The standard deviation of the height fluctuations is $3.8\unit{nm}$ (Fig.~S3a) . Histogram of the height fluctuations is close to a Gaussian distribution and the power spectrum of the height fluctuations is close to a Lorentzian. From the fitting, we obtained the characteristic wavelength to be $\lambda\sim0.75\unit{\mu m}$. External mechanical perturbations to the moving particle on a rough substrate can be estimated from potential energy calculations. The potential energy fluctuations calculated by assuming that the particle is moving along the ITO surface landscape are less or of the same order of the thermal fluctuation $k_{\mathrm{B}}T$. However, surface roughness of the order of $4\unit{nm}$ is close to the size of Debye length of the surface double layer of the particle. This Debye length level of roughness may cause even stronger fluctuations in the flow of charge around the particle and its distribution, which may have a significant impact on the effective rotational diffusion of the particle. Since the mean speed of the particles averaged over the period of the externally applied electric field is about $v_{\mathrm{m}}\sim60\unit{\mu m/s}$ and the characteristic time of the Lorentzian in Fig.3b is $\tau\sim0.032\unit{s}$, the number of fluctuation events experienced by a particle within $\tau$ is estimated by $n\sim v_{\mathrm{m}}\tau/\lambda$. We obtained $n\sim 2.4$. This number of fluctuation events is sufficient to cause loss of orientation information and result in athermal rotational diffusion of the particle.

\begin{figure}[h]
\begin{center}
\includegraphics[width=15.1cm]{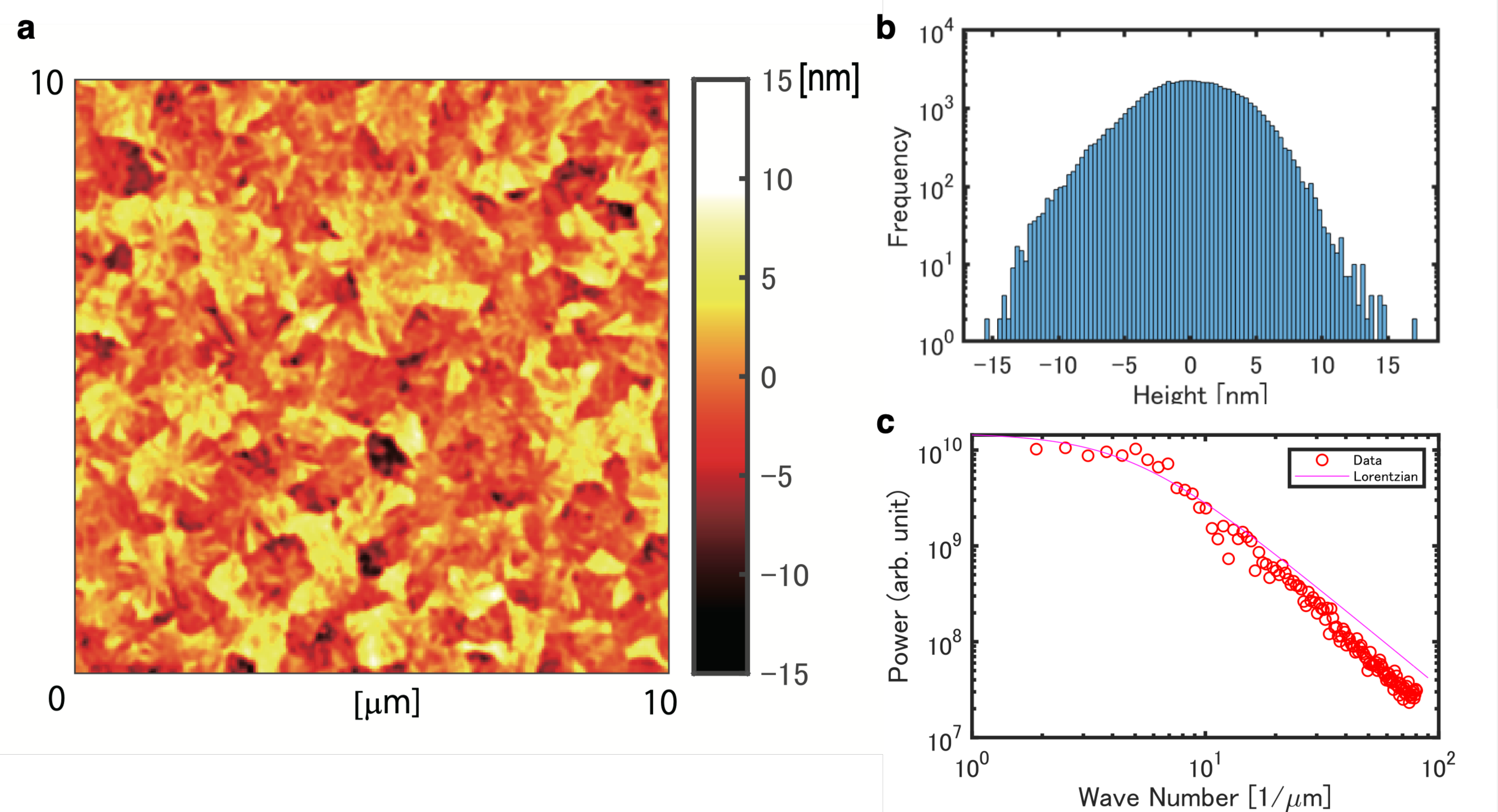}
\caption{Measurement of surface roughness of ITO substrate. (a) Height roughness of the ITO coat on the glass substrate measured with an AFM ($xy$-area size: $10\times10\unit{\mu m}^2$).  $Z$-range of color-coded image is $32.7 \unit{nm}$. (b) Histogram of height fluctuations. Standard deviation of the height fluctuation is $3.8\unit{nm}$. (c) Radial distribution of the height fluctuation power spectrum in $k$-space calculated from the two-dimensional Fourier transformation of the height fluctuations. Lorentzian fit gives a characteristic wavenumber of $4.8 \unit{\mu m}^{-1}$. (wavelength $\sim 0.75 \unit{\mu m}$)}
\label{fig:sup_surface}
\end{center}
\end{figure}

\newpage
\section*{Supplementary videos}
\begin{itemize}
    \item {\bf Video S1 (S1.mov): Short-time behavior of an isolated particle.} The particle reciprocates at the external frequency ($300\unit{Hz}$), though it is only barely visible in the movie. The movie is played at $0.01$ times the real speed.
    \item {\bf Video S2 (S2.mov): Long-time behavior of an isolated particle (stroboscopic).} This movie shows long-time persistent motion of the particle. The frame acquisition rate is set to be equal to the external frequency ($300\unit{Hz}$). The movie is played at $0.1$ times the real speed.
    \item {\bf Video S3 (S3.mov): A wide-field video (x10 objective) showing multiple clusters (stroboscopic, $250\unit{Hz}$).} Clusters are formed and maintained longer than $1000$ periods if they do not encounter other particles. The movie is played at $0.1$ times the real speed.
    \item {\bf Video S4 (S4.mov): Short-time behavior of a doublet ($300\unit{Hz}$).} The two particles reciprocate at the external frequency  in a cooperative manner. The propelling directions tend to align, but not always. The movie is played at $0.01$ times the real speed.
    \item {\bf Video S5 (S5.mov): Long-time behavior of a doublet (stroboscopic, $300\unit{Hz}$).} The two particles show persistent motions, with the interparticle distance fluctuating but kept in some range. The movie is played at $0.1$ times the real speed.
    \item {\bf Video S6 (S6.mov): Short-time behavior of a doublet ($250\unit{Hz}$).} The two particles reciprocate at the external frequency  in a cooperative manner. The propelling directions tend to align, but not always. The movie is played at $0.01$ times the real speed.
    \item {\bf Video S7 (S7.mov): Long-time behavior of a doublet (stroboscopic, $250\unit{Hz}$).} The two particles show persistent motions, with the interparticle distance fluctuating but kept in some range. The movie is played at $0.1$ times the real speed.
    \item {\bf Video S8 (S8.mov): Short-time behavior of a triplet ($300\unit{Hz}$).} The three particles reciprocate at the external frequency ($300\unit{Hz}$) in a cooperative manner. The propelling directions tend to align, but not always. The movie is played at $0.01$ times the real speed.
    \item {\bf Video S9 (S9.mov): Long-time behavior of a triplet (stroboscopic, $300\unit{Hz}$).} The three particles show persistent motions, with the interparticle distances fluctuating but kept in some range. The movie is played at $0.1$ times the real speed.    \item {\bf Video S10 (S10.mov): Short-time behavior of a triplet ($250\unit{Hz}$).} The three particles reciprocate at the external frequency in a cooperative manner. The propelling directions tend to align, but not always. The movie is played at $0.01$ times the real speed.
    \item {\bf Video S11 (S11.mov): Long-time behavior of a triplet (stroboscopic, $250\unit{Hz}$).} The three particles show persistent motions, with the interparticle distances fluctuating but kept in some range. The movie is played at $0.1$ times the real speed.
    \item {\bf Video S12 (S12.mov): Long-time behavior of a triplet formation (stroboscopic, $300\unit{Hz}$).} An initially isolated particle joined a doublet and formed a triplet. The shape of the triangle formed by the three particles dynamically changes. The movie is played at $0.1$ times the real speed.

\end{itemize}

%\bibliographystyle{rsc} %
%\bibliography{sample}